\documentclass[prd,twocolumn,10pt,aps,longbibliography,nofootinbib,superscriptaddress]{revtex4-2}

\usepackage{amsmath,amssymb,amsthm}
\usepackage{graphicx}
\graphicspath{{../figures/}}
\usepackage{tikz}
\usepackage{comment}

\usepackage{color}
\definecolor{darkblue}{rgb}{0,0,.5}
\definecolor{BLUE}{rgb}{0,0,1}
\definecolor{BLACK}{rgb}{0,0,0}
\usepackage[linktocpage, colorlinks=true, linkcolor=darkblue, citecolor=darkblue]{hyperref}
\usepackage{xr}
\usepackage{cancel}

\newcommand{\C}[1]{{\cal{#1}}}
\newcommand{\bb}[1]{\textbf{#1}}
\newcommand{\bs}[1]{\boldsymbol{#1}}

\newcommand{\lr}[1]{{\langle {#1}\rangle}}

\newcommand{\rl}[0]{{\rangle\langle}}

\newcommand{\tm}[0]{\text{$-$}}

\usepackage{scalerel}
\usetikzlibrary{calc}
\usetikzlibrary{patterns}
\usetikzlibrary{svg.path}
\definecolor{orcidlogocol}{HTML}{A6CE39}
\tikzset{
	orcidlogo/.pic={
		\fill[orcidlogocol] svg{M256,128c0,70.7-57.3,128-128,128C57.3,256,0,198.7,0,128C0,57.3,57.3,0,128,0C198.7,0,256,57.3,256,128z};
		\fill[white] svg{M86.3,186.2H70.9V79.1h15.4v48.4V186.2z}
		svg{M108.9,79.1h41.6c39.6,0,57,28.3,57,53.6c0,27.5-21.5,53.6-56.8,53.6h-41.8V79.1z M124.3,172.4h24.5c34.9,0,42.9-26.5,42.9-39.7c0-21.5-13.7-39.7-43.7-39.7h-23.7V172.4z}
		svg{M88.7,56.8c0,5.5-4.5,10.1-10.1,10.1c-5.6,0-10.1-4.6-10.1-10.1c0-5.6,4.5-10.1,10.1-10.1C84.2,46.7,88.7,51.3,88.7,56.8z};
	}
}
\newcommand\orcid[1]{\!%
	\href{https://orcid.org/#1}{%
		\mbox{%
			\scaleto{%
				\begin{tikzpicture}[yscale=-1,transform shape]
				\pic{orcidlogo};
				\end{tikzpicture}
			}{8pt}%
		}%
	}%
}

\begin{document}

\newtheorem{mydef}{Def.}[section]
\newtheorem{lemma}{Lemma}
\newtheorem{conj}{Conjecture}
\newtheorem{thm}{Theorem}
\newtheorem*{thm*}{Theorem}
\newtheorem{res}{Result}[section]
\theoremstyle{remark}
\newtheorem{rmrk}{Remark}[section]

\title{Redundant Records of the Past: Unifying Quantum Darwinism and Decoherent Histories}

\author{Philipp Strasberg~\orcid{0000-0001-5053-2214}}
\affiliation{Instituto de F\'isica de Cantabria (IFCA), Universidad de Cantabria--CSIC, 39005 Santander, Spain}

\begin{abstract}
 There exist two different paradigms for records in isolated quantum systems: decoherent histories (which determines which past events leave formal records) and quantum Darwinism (which determines which current events get redundantly recorded). Neither one implies the other, and connections between the two are scarcely investigated. We unify quantum Darwinism and decoherent histories based on the insight that only redundant records about \emph{past} events matter. The resulting framework of \emph{past redundancy} encompasses quantum Darwinism when applied to events at the current time, and it implies the probability sum rule or decoherence of histories. The Bhattacharyya coefficient (or fidelity) plays an important role to establish this connection. Importantly, various examples show that past redundancy is a more widespread phenomenon than the fine-tuned models necessary for quantum Darwinism suggest. We thus arrive at a richer picture in which redundancy can be caused by mechanisms different from quantum Darwinism (such as macroevents or cascade structures), which emphasizes that redundancy is necessary but not sufficient for objectivity, and which, overall, allows to realistically investigate the question: how generic is redundancy in nature?
\end{abstract}

\maketitle

\section{Introduction}

Quantum paradoxes arise from trying to understand the relation between the observer and the observed. Thus, an important question is: what can actually be observed, i.e., which events in a quantum system leave records accessible to observers? And how does one answer this question for a unitarily evolving, isolated quantum systems without invoking the \emph{a priori} existence of a classical world as in the Copenhagen interpretation? Two important milestones of progress are the following.

The framework of decoherent histories (DH)~\cite{OmnesRMP1992, HartleLecture1992, GellMannHartlePRD1993, HalliwellANY1995, DowkerKentJSP1996} expresses quantum evolution as a sum over histories similar to Feynman's decomposition into an integral over paths (in fact, a path is the continuous and fine-grained limit of a history). Its main purpose is to \emph{decide}, via the so-called DH condition, which events during the evolution of a quantum system can be unambiguously discriminated in the final wave function $|\Psi_f\rangle$. For a globally pure state (which will be our interest here) this is an \emph{if and only if} condition: observers can only have records about past events that form DH---a crucial insight that quickly dissolves~\cite{LosadaLauraLombardiPRA2019} apparent paradoxes in quantum mechanics~\cite{FrauchigerRennerNC2018}.

Second, there is the framework of quantum Darwinism (QD)~\cite{ZurekNP2009, KorbiczQuantum2021, ZurekEnt2022}. It provides a \emph{mechanism} for how the current information about certain preferred ``pointer'' states of an open quantum system proliferates and becomes \emph{redundantly encoded} in many fragments of its environment.

There is a natural tension between these two frameworks. While the DH formalism solely decides which past events leave \emph{formal records} without asking how they can be accessed (but see Ref.~\cite{HalliwellPRD1999} for an exceptional model-specific study), QD focuses on the practical utility of records of the \emph{current} state of the system. Neither one of the two frameworks implies the other. By unifying the ideas of DH and QD, we introduce a framework that explains the emergence of locally accessible, redundant records of (multiple events in) the past.

\emph{En passant}, this framework solves another limitation of QD, namely that QD relies on specific, fine-tuned (non-generic) system-environment ($SE$) models. In fact, if the system and environment are composed out of qubits, the requirement of perfect QD enforces an almost unique operator structure on the $SE$ Hamiltonian $H_{SE} = H_S + H_E + V_{SE}$~\cite{DuruisseauTouilDeffnerEntropy2023}, and relaxing the qubit assumption still entails an extensive amount of commutator constraints on $H_{SE}$~\cite{DoucetDeffnerPRX2024} (see also Ref.~\cite{SchwarzhansEtAlPRR2025}). Essentially, QD is incompatible with modern ideas of quantum thermalization such as volume law entanglement~\cite{BianchiEtAlPRXQ2022} and the eigenstate thermalization hypothesis~\cite{DeutschRPP2018}---in contrast to DH which are well compatible with them~\cite{StrasbergReinhardSchindlerPRX2024, WangStrasbergPRL2025}. This causes a conundrum as it implies that QD and thermalization can not both be generic~\cite{LeWinterAdessoEntropy2021}. Motivated by this shortcoming, a recent example suggests that QD can arise also in generic, chaotic environments, but only if $V_{SE}$ and $H_E$ can be switched off/on and if $V_{SE}$ induces a macroscopic energy change in the environment~\cite{CaoNussinovArXiv2026}.

In contrast, we will find that, once freed from the idea that the records have to be redundantly correlated with the \emph{current} system state, it follows naturally that, first, a larger class of models can generate past redundancy and, second, mechanisms different from QD (which Zurek describes as an ``environment [...] relentlessly cloning the information about the einselected pointer states''~\cite{ZurekRMP2003}) can also cause redundancy. This provides an important step forward towards answering a question that initially motivated this project: \emph{how generic is redundancy?}

Remarkably, attempts to combine DH and QD are scarce. There are two very recent case studies~\cite{FerteFarciCaoPRL2026, ArefyevaPolyakovPRA2026} and a general attempt by Riedel \emph{et al.}~\cite{RiedelZurekZwolakPRA2016} to formally combine DH and QD using Finkelstein's notion of partial trace decoherence~\cite{FinkelsteinPRD1993}. However, while the connection between DH and partial trace decoherence was investigated in detail, Riedel \emph{et al.} announced that ``the connection to QD will be completed in a forthcoming companion paper''~\cite{RiedelZurekZwolakPRA2016}, which has never been published. A preliminary draft of this paper---shared with the author after completion of this work~\cite{RiedelPC2026}---argues (in unison with some of our ideas below) that the notion of an open system with a fixed, preferred pointer basis is not necessary to define redundancy (see also Ref.~\cite{RiedelPRL2017}).

\emph{Overview.} In Sec.~\ref{sec:QD} we introduce QD and three definitions of redundancy that all rely on the single-time correlations in a $SE$ state $\rho_{SE}(t)$. We also discuss four open questions: no account of past redundancy, fine-tuned Hamiltonians, fragility of the information in the environment, and fine-tuned initial states. This paper claims to solve at least the first two questions, and to this end Sec.~\ref{sec:basics} introduces the basics of past redundancy and clarifies its relation to QD when applied to current events. Importantly, when applied to past events, we will see in Sec.~\ref{sec:examples} for various examples that past redundancy is more widespread than QD and does not oppose thermalization. In Sec.~\ref{sec:extensions} we treat more complicated situations (correlated records and multiple events) and show that redundancy alone implies classicality. Conclusions are drawn in Sec.~\ref{sec:conclusions}.

\section{Quantum Darwinism: Short Review and Open Questions}\label{sec:QD}

In a letter to Einstein Schr\"odinger mentions in 1950 (my italics): ``The conception of a world that \emph{really exists} is based on there being a far-reaching \emph{common experience} of many individuals, in fact of all individuals who come into the same or a similar situation with respect to the object concerned.'' This common experience requires that events leave redundant records such that different observers, having access to \emph{a priori} different information (i.e., different parts of the world), can come to an agreement about what happened.

It took 50 years until the first formal attempt to define redundant records was made by Zurek~\cite{ZurekAP2000}, which culminated in the framework of QD~\cite{ZurekNP2009, KorbiczQuantum2021, ZurekEnt2022}. It is important to remark here that, in principle, one should distinguish between the \emph{concept} of redundancy (a specific property that a quantum system may have or not) and the mechanism that \emph{causes} redundancy (which is QD). Since QD is so far the only attempt to explain redundancy within quantum mechanics, these notions are often used synonymously. We will try to be more careful with the terminology since we will see that redundancy $\neq$ QD.

\subsection{General Formalism}

\begin{figure}[t]
	\centering\includegraphics[width=0.35\textwidth,clip=true]{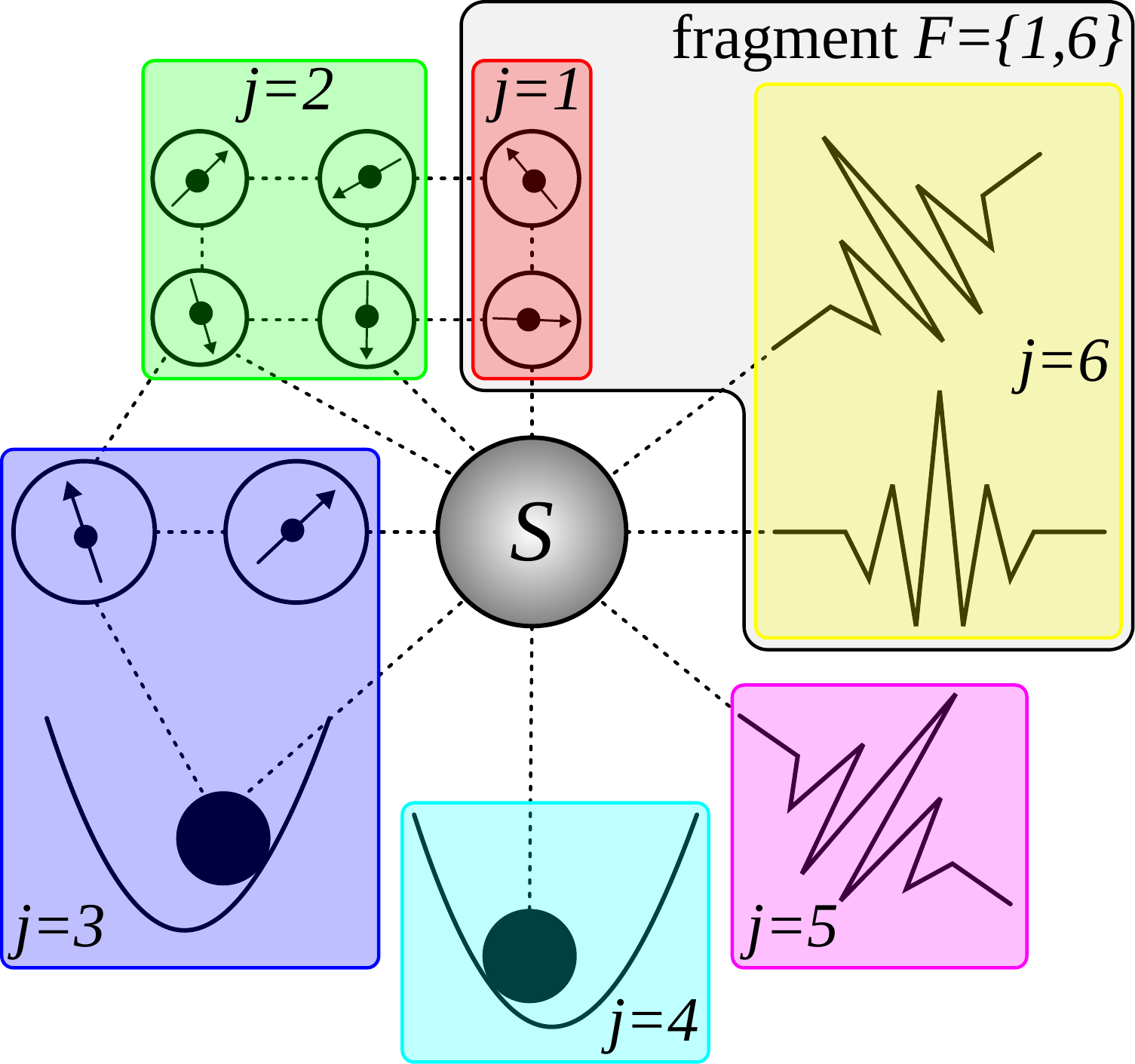}
	\caption{A system $S$ coupled to an environment composed of many different degrees of freedoms (spins, oscillators, photons, etc.). The question is whether the environment can be split in a natural way into $N$ accessible, elementary fragments $j$ (here, $N=6$) such that they encode redundant information about the system as quantified, e.g., by eqn~(\ref{eq:QD:MI}). In general, a fragment $F$ can be composed of multiple elementary fragments (here we sketched $F=\{1,6\}$).}
	\label{fig:QD setup}
\end{figure}

Central to QD is a system $S$ coupled to an environment $E$ split into $N$ fragments $j$, i.e., we assume a global tensor product structure of the form $\C H_S\otimes\C H_E$ with $\C H_E = \bigotimes_{j=1}^N \C H_j$. Note that there is no further requirement on $\C H_j$: all the $\C H_j$ could be different and they could describe composite systems themselves. Moreover, we do not ask where this tensor product structure comes from~\cite{CarrollSinghPRA2021, LoizeauSelsFP2025, SoulasFranzmannDiBiagioArXiv2025} and assume it as naturally given (e.g., by locality considerations). We call each $j$ an ``elementary'' fragment, but in general we allow a fragment $F = \{j_1,\dots,j_n\}$ of size $n=|F|$ to be a composite of $n$ elementary fragments (with $n=1$, a single elementary fragment, and $n=N$, the entire environment, as limiting cases). Furthermore, $\overline F\supset F$ denotes an extended fragment and $F' = \{1,\dots,N\}\setminus F$ the complementary fragment. See Fig.~\ref{fig:QD setup} for a sketch. The quantum mutual information (MI) between the system $S$ and a fragment $F$ is $I_Q(S:F) = S_S+S_F-S_{SF}$ with $S_X = -\mbox{tr}\{\rho_X\ln\rho_X\}$ the von Neumann entropy of the reduced density matrix $\rho_X$ with $X \in \{S,F,SF\}$. The joint $SE$ state $|\Psi_t\rangle = U_{t,s}|\Psi_s\rangle$ is propagated from time $s$ to $t$ by a unitary time-evolution operator $U_{t,s}$ generated by a (possibly time-dependent) $SE$ Hamiltonian. We now review three related but non-equivalent definitions of redundancy.

First, redundancy was defined by considering the quantum MI $I_Q(S:fE)$ between the system $S$ and a fraction $f\in(0,1)$ of the environment $E$, i.e., a fragment $F$ of size $|F| = fN$, and by positing that~\cite{ZurekNP2009}
\begin{equation}\label{eq:QD:MI}
 I_Q(S:fE) \approx S_S,
\end{equation}
holds for a large range of fractions $0<f<1$, giving rise to a characteristic \emph{redundancy plateau}.

It has been noted, however, that $I_Q(S:fE)$ can be dominated by quantum discord instead of classical correlations and therefore it can contain information that is classically inaccessible~\cite{HorodeckiKorbiczHorodeckiPRA2015, PleasanceGarrawayPRA2017, LeOlayaCastroPRA2018, KorbiczQuantum2021}. It was instead posited that eqn~(\ref{eq:QD:MI}) should be replaced with a \emph{spectrum broadcast structure} (SBS) of the joint state $\rho_{SfE}$ of the system and a large environmental fraction $fE$, i.e., for a sufficiently large $f<1$ one posits that
\begin{equation}\label{eq:QD:SBS}
 \rho_{SfE} = \sum_ x \sqrt{p(x)} \rho_S(x) \bigotimes_{j\in fE} \rho_{j}(x).
\end{equation}
Here, the different normalized states have orthogonal support, i.e., for $x\neq x'$ we have $\rho_S(x) \perp \rho_S(x')$ or $\rho_j(x) \perp \rho_j(x')$, and $p(x)$ is a probability distribution. Equation~(\ref{eq:QD:SBS}) implies eqn~(\ref{eq:QD:MI}), but the converse requires further conditions~\cite{LeOlayaCastroPRL2019, FellerEtAlPRL2021, LeOlayaCastroPRL2021}.

It is worth to note that, when viewing QD as a refinement of environmentally-induced decoherence, one expects that the $\rho_S(x)$ commute with the ``einselected'' pointer basis of the system~\cite{ZurekRMP2003}. The idea that there must be a unique pointer basis is often demanded as an additional criterion for perfect QD. The existence of a preferred pointer basis plays, however, no decisive role in our generalized formalism below. We will come back to this subtlety in context of the example of Sec.~\ref{sec:coherent state}.

A third definition of redundancy works directly with the wave function of the isolated systems, where one posits a so-called \emph{branching form}~\cite{ZurekEnt2022, TouilEtAlQuantum2024}:
\begin{equation}\label{eq:QD:branching form}
 |\Psi\rangle_{SE} = \sum_x \sqrt{p(x)} |x\rangle_S \bigotimes_{j=1}^N |\varphi_j(x)\rangle.
\end{equation}
Here, $|x\rangle$ is said pointer basis and the $|\varphi_j(x)\rangle\in\C H_j$ are linearly independent but \emph{not} necessarily orthogonal. However, approximate orthogonality is quickly restored when considering a fragment $F_\alpha$ of size $n\gg 1$. Thus, if one decomposes the environment into such sufficiently large fragments, $E = \bigcup_\alpha F_\alpha$, a reduced state of SBS emerges for those fragments whenever one traces out one such fragment, i.e., we effectively have to replace the label $j$ by $\alpha$ in eqn~(\ref{eq:QD:SBS}). Note the tacit assumption here that one can simultaneously satisfy $n\gg 1$ (to get orthogonality) and $N\gg n$ (to get the redundancy plateau).

\subsection{Generalized CNOT-fan-out models}

\begin{figure}[t]
	\centering\includegraphics[width=0.40\textwidth,clip=true]{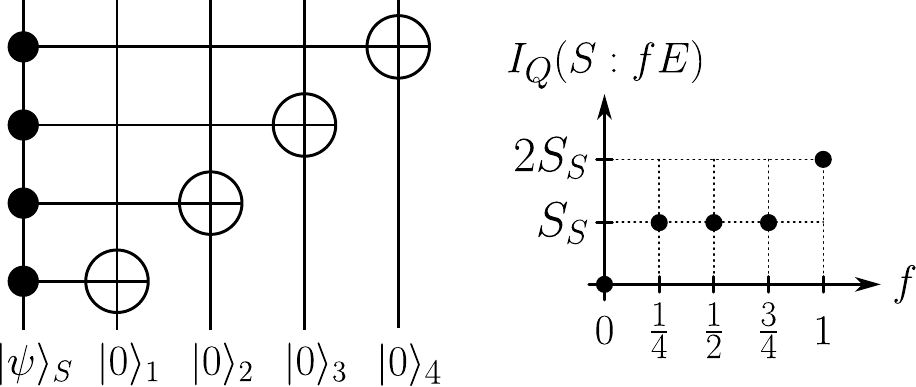}
	\caption{A simple CNOT-fan-out quantum circuit. A system $S$ controls a NOT operation in each of the four qubits comprising different fragments of a non-interacting environment. For instance, for the initial system state $c_0|0\rangle_S+c_1|1\rangle_S$ the resulting final state is $c_0|00000\rangle + c_1|11111\rangle$ (with $|00000\rangle \equiv |0\rangle_S\otimes|0\rangle_1\otimes|0\rangle_2\otimes|0\rangle_3\otimes|0\rangle_4$ and similarly for $|11111\rangle$). To the right: the plot of the quantum MI after the circuit has been applied shows the characteristic redundancy plateau.}
	\label{fig:cnot circuit}
\end{figure}

The core idea of QD as a mechanism to produce redundancy can be explained by a suitable generalization of a CNOT-fan-out quantum circuit (an example of which is shown in Fig.~\ref{fig:cnot circuit}), which covers many model specific studies in the literature. The resulting unitary of such a generalized CNOT-fan-out quantum circuit is given by
\begin{equation}\label{eq:U CNOT-fan-out}
 U = \sum_x \Pi_S(x) \bigotimes_j U_j(x).
\end{equation}
Here, $\{\Pi_S(x)\}$ is a complete set of orthogonal system projectors, which can be identified with the pointer states of the system. The unitaries $U_j(x)$ acting on fragment $j$ need to satisfy $U_j(x)\neq U_j(x')$ for $x\neq x'$. Note that, due to the trivial system and environment dynamics, it does not matter in which order the different fragments $j$ interact with the system in a quantum circuit. They could also interact all at once with the system via an interaction Hamiltonian $V_{SE} = A_S\otimes\sum_j B_j$, where $A_S$ is a system operator commuting with all $\Pi_S(x)$ and $B_j$ acts non-trivially only on fragment $F_j$.

When applied to an initial state of product form
\begin{equation}\label{eq:initial product state}
 |\Psi(0)\rangle = |\psi(0)\rangle_S\bigotimes_j |\varphi\rangle_j, ~~~ |\psi(0)\rangle = \sum_x \sqrt{p(x)} |x\rangle_S
\end{equation}
with $\sqrt{p(x)} |x\rangle_S = \Pi_S(x)|\psi(0)\rangle$, the resulting final state is
\begin{equation}
 U|\Psi(0)\rangle = \sum_x \sqrt{p(x)} |x\rangle_S \bigotimes_j |\varphi_j(x)\rangle
\end{equation}
with $|\varphi(x)\rangle_j \equiv U_j(x)|\varphi\rangle_j$. This state has branching form, and thus generates redundancy, whenever $\lr{\varphi(x)|\varphi(x')}_j \neq 1$ for all $x\neq x'$ and sufficiently many fragments $j$.

\subsection{Open questions}

\subsubsection{No past redundancy}\label{sec:no past redundancy}

All definitions above are expressed in terms of \emph{single-time} quantities to describe \emph{spacelike} correlations in a quantum state: $\rho_{SE}(t)$ alone decides about redundancy without reference to the dynamics or any space\emph{time} correlations in two-time (or higher) correlation functions.

However, already the word ``record'' implies an implicit reference to two points in time. For instance, the redundantly encoded information in the cosmic microwave background is a fingerprint of the Universe' state during the recombination epoch, around 13.5 billion years ago, but the Universe no longer looks like during the recombination epoch. Likewise, dinosaur fossils correlate with the fact that there were dinosaurs and not that there are dinosaurs. Also every quantum experiment we performed refers to states of quantum systems in the past. Last but not least, even if we see an object ``now'', we actually see its state a time $d/c$ ago (with $d$ the distance to that object and $c$ the speed of light).

While we will see that a generalized CNOT-fan-out quantum circuit also creates past redundancy (rather trivially), none of the definitions~(\ref{eq:QD:MI}), (\ref{eq:QD:SBS}) or~(\ref{eq:QD:branching form}) implies the formation of records about the past. Importantly, as we will demonstrate with various examples below, a larger class of situations can generate past redundancy but not QD. This brings us directly to the next point.

\subsubsection{Fine-tuned unitary evolution}

While the generalized CNOT-fan-out unitary operator in eqn~(\ref{eq:U CNOT-fan-out}) is a popular theoretical model, very few \emph{real} physical systems (i.e., systems that are not artificially engineered) actually obey it. In fact, one reason to award the Nobel prize to Serge Haroche was that his lab was able to engineer a quantum non-demolition measurement that obeys eqn~(\ref{eq:U CNOT-fan-out})~\cite{HarocheRMP2013}---where the system $S$ is a microwave cavity and the fragments $j$ are atoms flying through the cavity. We will study this setup in Sec.~\ref{sec:Haroche} but even there deviations from eqn~(\ref{eq:U CNOT-fan-out}) are relevant. Essentially, a quantum system described by eqn~(\ref{eq:U CNOT-fan-out}) is easily solvable, but we know that most quantum systems are not. Note in particular that eqn~(\ref{eq:U CNOT-fan-out}) does not lead to thermalization. While one can relax eqn~(\ref{eq:U CNOT-fan-out}), the requirement of perfect QD still imposes an extensive number of constraints~\cite{DuruisseauTouilDeffnerEntropy2023, DoucetDeffnerPRX2024} and many studies found a sensitive dependence on the structure of the bath~\cite{RiedelZurekZwolakNJP2012, GiorgiGalveZambriniPRA2015, PleasanceGarrawayPRA2017, MilazzoEtAlPRA2019, OliveiraPaulaDrumondPRA2019, CampbellEtAlPRA2019, ValladaresDuqueStrasbergEPL2025}.

There is, however, one important environment which is often cited as the prime and most important example for redundancy and QD: scattered photons~\cite{RiedelZurekPRL2010, RiedelZurekNJP2011, KorbiczHorodeckiHorodeckiPRL2014}. We will finish this paper with a critical discussion of this case in Sec.~\ref{sec:conclusions:photons}. For now it suffices to say that there are stable and redundant records about past events (e.g., dinosaur bones or outcomes of quantum experiments) that are independent of any photonic environment and that also demand understanding.

Since the full class of models that generates approximate QD is currently not known, it is beyond doubts desirable to have a larger class of models capable of generating redundant records. See also Refs.~\cite{CaoNussinovArXiv2026, SchneeweissArXiv2026} for recent attempts to move beyond toy models.

\subsubsection{Fragility of the records}

One reason to favor a definition in terms of the SBS over the definition given in eqn~(\ref{eq:QD:MI}) is that the quantum MI quantifies the totality of possible correlations, independent of the question whether they are locally accessible~\cite{HorodeckiKorbiczHorodeckiPRA2015, KorbiczQuantum2021}. This criticism can be extended to the SBS or branching form because they still assume that \emph{every} local observable in $\C H_j$, as defined by the tensor product decomposition $\C H_E = \bigotimes_j \C H_j$ of the environmental Hamiltonian, is \emph{accessible}. However, human and other observers are limited even in what they can measure locally. This is discussed in further detail for photonic environments in Sec.~\ref{sec:conclusions:photons}.

The criticism can be strengthened by pointing out that, when the unitary in eqn~(\ref{eq:U CNOT-fan-out}) is generated by a Hamiltonian, the $U_j(x)$ will depend on time. Hence, also the $\rho_j(x)$ and the $|\varphi_j(x)\rangle$ in the SBS and branching form, respectively, will acquire a time-dependence. This (typically very fast) time-dependence plays no role in previous definitions of redundancy. While the pointer states of the system are fixed, stable and time-independent, the records in the environment are not. Redundancy does not imply easily accessible redundancy or, put differently, redundancy is necessary but not sufficient for objectivity.

We can not claim that past redundancy solves this problem immediately. However, it raises awareness of it because we will be forced to define past redundancy with respect to a particular local measurement basis.

\subsubsection{Fine-tuned initial states}

Apart from the dynamics, also the initial state in eqn~(\ref{eq:initial product state}) is fine-tuned as it demands vanishing correlations between an extensive number of subsystems and model-specific studies of QD focus on these initial states. As Doucet and Deffner write~\cite{DoucetDeffnerPRX2024}: a ``constraint necessary for a model to exhibit QD [is that] the initial system-environment state must not possess quantum correlations in the environment.'' Yet, generic states of multipartite systems are highly entangled~\cite{PagePRL1993, ZyczkowskiEtAlPRA1998, PopescuShortWinterNatPhys2006, BianchiEtAlPRXQ2022}.

However, if QD is fundamentally about explaining the emergence of redundancy in a \emph{quantum} Universe, it seems legitimate to take advantage of the power of the Multiverse. Any initial state of the environment can be decomposed as a sum of decorrelated states,
\begin{equation}\label{eq:initial state decomp}
 |\psi_0\rangle_E = \sum_{\alpha=1}^r \Gamma_\alpha |\alpha\rangle \text{ with } |\alpha\rangle = |\alpha_1\rangle\otimes\cdots\otimes|\alpha_N\rangle,
\end{equation}
where $r$ is the product of Schmidt ranks associated with a successive Schmidt decomposition of the multipartite environmental state  ($r$ is not unique and its smallest possible value has been suggested as a good measure of multipartite entanglement~\cite{EisertBriegelPRA2001}). Note that eqn~(\ref{eq:initial state decomp}) is an orthogonal decomposition (i.e., $\lr{\alpha|\alpha'} = \delta_{\alpha,\alpha'}$) even though the local vectors in general do \emph{not} obey $\lr{\alpha_j|\alpha'_j} = \delta_{\alpha_j,\alpha'_j}$. Each vector in eqn~(\ref{eq:initial state decomp}) would then give rise to a branch in which the same system information is redundantly encoded, only how it is encoded, i.e., the $|\varphi_j(x)\rangle$ from eqn~(\ref{eq:QD:branching form}), differs from branch to branch. Hence, the branches $|\varphi_j(x,\alpha)\rangle$ would depend on $\alpha$. If local observers could distinguish different $\alpha$, i.e., if $\lr{\varphi(x,\alpha)|\varphi(x',\alpha')}_j\neq 1$ for all $j$ and all $(x,\alpha)\neq(x',\alpha')$, then QD would follow again.

Since the focus of this paper is on a formulation of past redundancy encompassing QD, we will not go into further details here and in all the examples we will use product states for simplicity. For future applications it might be useful to keep this extension in mind as it is possible to include information about the initial environmental state $\alpha$ into the formalism of past redundancy.

\subsubsection{Brief summary}

We have identified four open questions in QD of which (to the best of my knowledge) the first and third have not yet received any attention. We believe it is necessary to answer those questions before one can claim that one has understood the redundancy of records and the emergence of objectivity. The framework of past redundancy we will propose now, which is broader than QD, is designed to solve the first question and, in doing so, allows clear progress for the second. We will not attempt a general solution to the third question here, but the concept of past redundancy might provide the right tools to address it: it indeed forces us to think about this question for the examples of Sec.~\ref{sec:examples}, also see Sec.~\ref{sec:conclusions:photons}. An answer to the fourth question has been outlined above, and we will also see in Sec.~\ref{sec:dependency} that there is a broader class of states than those in eqns~(\ref{eq:QD:SBS}) and~(\ref{eq:QD:branching form}) that show redundancy.

\section{Basics of past redundancy}\label{sec:basics}

We here introduce the basic concept of past redundancy for a single event in the past. Even though the framework can be seen as an application of the DH formalism, no familiarity with it is required for now. Instead, we aim at a minimal version of past redundancy that (a) is sufficient to understand the examples in Sec.~\ref{sec:examples} and (b) makes the connection to QD transparent. Extensions are treated in Sec.~\ref{sec:extensions}.

The overall setup of a system coupled to an environment split into many fragments with respect to some preferred tensor product decomposition, and altogether evolving unitarily, is the \emph{same} as in QD. Moreover, we introduce the classical MI $I_C(X:Y) = H(X) + H(Y) - H(XY)$ between two random variables $X$ and $Y$ with $H(A)$ the Shannon entropy of a random variable $A\in\{X,Y,XY\}$. Crucial for our application below is the fact that the classical MI can be defined for random variables at different times, e.g., $X_sY_t$ given a joint distribution $p(x_s,y_t)$, whereas there is no obvious candidate for a two-time quantum MI owing to the fact that there is no obvious candidate for a two-time density matrix $\rho(s,t)$ (however, proposals exist~\cite{TarantoEtAlArXiv2025}).

\subsection{Past events}\label{sec:past_events}

We first need to specify what we mean by an event in the past. To this end, consider $|\psi'_t(x_s)\rangle \equiv U_{t,s}\Pi_S(x)U_{s,0}|\Psi_0\rangle$ with $\Pi_S(x)$ some projector describing some physical property $x$ of the open system (e.g., the state of a spin such as $x=$ $\uparrow_z$), $U_{t,s}$ the unitary time evolution operator and $|\Psi_0\rangle$ the initial wave function of the system and environment. The wave function $|\psi'_t(x_s)\rangle$ describes the unitarily evolved state of an isolated quantum systems at time $t$ \emph{conditioned} on passing through some Hilbert subspace $\C H_x$, as defined by the projector $\Pi_S(x)$, at time $s$. We will interpret $t$ as the current time (``$t_\text{now}$'') and we call properties, such as $x$, that a quantum system could have had at an earlier time $s$ an \emph{event} $(x;s)$. Note that $|\psi'_t(x_s)\rangle$ is not normalized, and therefore we used a prime in the notation. The normalized state $|\psi_t(x_s)\rangle = |\psi'_t(x_s)\rangle/\|\psi'_t(x_s)\|$ is denoted without prime.

In general, a quantum system will be in a superposition of different events $x$ at time $s$ (such as $x\in\{\uparrow_z,\downarrow_z\}$ for a spin-1/2 system), and therefore we consider a complete set of $m$ orthogonal projectors $\{\Pi_S(x)\}_{x=1}^m$. Owing to completeness, $\sum_{x=1}^m \Pi_S(x) = I_{\C H}$ (with $I_{\C H}$ the identity operator), we can decompose the unitary time evolution of the global wave function as
\begin{equation}\label{eq:mini-history}
 |\Psi_t\rangle = U_{t,0}|\Psi_0\rangle = \sum_x |\psi'_t(x_s)\rangle.
\end{equation}
As anticipated, $|\Psi_t\rangle$ is in a superposition of $|\psi'_t(x)\rangle$ for a complete set of possible events $(x;s)$. Note that, even though $|\psi_t(x)\rangle$ has an alternative interpretation as the state of a quantum system after a projective measurement at time $s$ gave outcome $x$, there is \emph{no} collapse of the wave function happening in eqn~(\ref{eq:mini-history}). Equation~(\ref{eq:mini-history}) is an exact rewriting of unitary time evolution.

We can now ask: can we find out at time $t$ which event $x$ happened at time $s$? \emph{A priori} the straightforward answer is: measure in the basis $\{\Pi_{t,s}(x)\}_{x=1}^m$ where $\Pi_{t,s}(x) = U_{t,s}\Pi_S(x)U^\dagger_{t,s}$ is the (backwards) Heisenberg evolved projector $\Pi_S(x)$. Since $\Pi_{t,s}(x)|\psi'_t(x'_s)\rangle = \delta_{x,x'}|\psi'_t(x_s)\rangle$, this discriminates with certainty the branches $|\psi'_t(x_s)\rangle$. However, there are three reasons why it is advisable to explore another route. First, this strategy does not generalize well beyond the simple single-event case (for which the DH condition becomes more useful, see Sec.~\ref{sec:extensions}). Second, $\Pi_{t,s}(x)$ will in general be a complicated, delocalized and time-dependent observable that is hard to measure as its support will no longer be restricted to the system. Third, $\Pi_{t,s}(x)$ does not help to understand the emergence of multiple, redundant records of the same event $(x;s)$, to which we turn now.

\subsection{Redundant records: A first definition}\label{sec:first_def}

We start with notation. Consider complete sets of orthogonal projectors $\{\Pi(r_j)\}_{r_j=1}^m$ acting locally on $\C H_j$ for every elementary fragment $j\in\{1,\dots,N\}$. For an arbitrary fragment $F=\{j_1,\dots,j_n\}$ we denote $r_F = (r_{j_1},\dots,r_{j_n})$ and $\Pi(r_F) = \Pi(r_{j_1})\otimes\cdots\otimes\Pi(r_{j_n})$. The (non-normalized) state conditioned on $r_F$ at time $t$ and $x$ at time $s$ is
\begin{equation}
 |\psi'(r_F,x_s)\rangle \equiv \Pi(r_F)|\psi'_t(x_s)\rangle.
\end{equation}
The probability to have event $x$ and record $r_F$ is given by its norm and can be written in different ways:
\begin{equation}
 \begin{split}\label{eq:prob_def}
  p(r_F,x) &= \lr{\psi'(r_F,x_s)|\psi'(r_F,x_s)} \\
  &= \lr{\Psi_s|\Pi(x)U_{t,s}^\dagger\Pi(r_F)U_{t,s}\Pi(x)|\Psi_s} \\
  &= \mbox{tr}\{\Pi(r_F)U_{t,s}\Pi(x)|\Psi_s\rl\Psi_s|\Pi(x)U_{t,s}^\dagger\}.
 \end{split}
\end{equation}
Note that we put no time subscripts $s$ and $t$ on $x$ and $r_F$, respectively, to avoid notational clutter. With these preliminary agreements we can define past redundancy.

We say that the projectors introduced above encode $N$ \emph{redundant records} with respect to the set of events $\{(x;s)\}_{x=1}^m$ if the following two conditions are satisfied.

\bb{(1)} The classical MI $I_C(X:R_F)$ exhibits a classical \emph{redundancy plateau}, i.e.,
\begin{equation}\label{eq:classical_plateau}
 I_C(X:R_F) \lesssim H(X)
\end{equation}
for all fragments $F$ that are sufficiently large but still much smaller than the total environment: $|F|\ll N$.
 
\bb{(2)} The probabilities $p(r_F,x_s)$ defined in eqn~(\ref{eq:prob_def}) can be interpreted \emph{classically}.

Some remarks are immediately in order. First, this is a \emph{qualitative} definition. During the further development of the theory below it will become clear that quantitative definitions are possible. However, the focus of this work is to introduce a novel conceptual framework, so it is advisable to not get lost in technical ``$\epsilon$-$\delta$-criteria''.

Second, $I_C(X:R_F)$ is monotonically increasing with fragment size. However, in contrast to quantum MI, we have $I_C(X:R_F)\le H(X)$ for all $F$. The redundancy plateau of classical MI thus looks different from QD, in particular for $|F|\ge N/2$, see Figure~\ref{fig PIP}($t_2$) for an example.

Third, we did not yet specify the meaning of ``classicality'' in the second point, which is related to the decoherence of histories and the probability sum rule. This will be clarified in Sec.~\ref{sec:classicality}. Moreover, it will later turn out that the redundancy plateau for classical MI \emph{implies} the classical behaviour of probabilities. This remarkable fact will be proven in Sec.~\ref{sec:extensions}. Keeping this in mind we can then see that the definition of past redundancy parallels the definition given in eqn~(\ref{eq:QD:MI}). Thus, we continue by making the connection to eqn~(\ref{eq:QD:MI}) rock solid.

\subsection{Connection to QD}\label{sec:QD_connection}

We here prove that, in the limiting case $s\rightarrow t$ where records and events are defined at the same time, the redundancy plateau for classical MI implies a redundancy plateau for quantum MI, i.e., eqn~(\ref{eq:classical_plateau}) implies eqn~(\ref{eq:QD:MI}). The converse is not true, and we will argue afterwards that this is a welcome feature as a way to resolve the debate between QD and the SBS.

Let $\Pi_S(x) = |x\rl x|_S$ be the projectors on the pointer basis. An elementary inequality of quantum information theory, $I_C(X:R_F) \le I_Q(S:F)$, provides us with a lower bound on the quantum MI. It remains to show that this lower bound is not too loose. To this end, suppose the opposite, i.e., the quantum MI notably exceeds the classically allowed value by a non-negligible $\delta>0$:
\begin{equation}\label{eq:assumption}
 I_Q(S:F) > S_S + \delta.
\end{equation}
By considering the quantum MI with respect to the complementary fragment $F' = E\setminus F$, we find
\begin{equation}
 I_Q(S:F') = 2S_S - I_Q(S:F),
\end{equation}
where we used $S_{F'} = S_{SF}$ and $S_{SF'} = S_F$ for a globally pure $SE$ state. It then follows from assumption~(\ref{eq:assumption}) that
\begin{equation}
 I_Q(S:F') \le S_S - \delta.
\end{equation}
But since $H(X) \ge S_S$, we then find by definition of the classical redundancy plateau
\begin{equation}
 S_S \le H(X) \approx I_C(X:R_{F'}) \le I_Q(S:F') \le S_S - \delta.
\end{equation}
This implies the desired contradiction $\delta\le 0$, i.e., eqn~(\ref{eq:assumption}) must be wrong.

It is not possible to establish the converse. In fact, the gap between quantum and classical MI is related to \emph{quantum discord}~\cite{OllivierZurekPRL2001, ModiEtAlRMP2012}, which does not need to vanish in presence of a redundancy plateau for quantum MI. This discrepancy was the reason to favor a definition of redundancy in terms of the SBS [eqn~(\ref{eq:QD:SBS})] over the original definition in eqn~(\ref{eq:QD:MI})~\cite{HorodeckiKorbiczHorodeckiPRA2015, PleasanceGarrawayPRA2017, LeOlayaCastroPRA2018, LeOlayaCastroPRL2019, KorbiczQuantum2021}. We can now anticipate that, in the equal-time limit $s\rightarrow t$, our definition sits \emph{in between}. By definition, classical MI guarantees that the redundant information is classically accessible, but the next section reveals that the condition of a SBS (or a branching form of the wave function) is too strong. 

\subsection{Conditional independence of the records}\label{sec:cond_indep}

We say the records satisfy conditional independence if 
\begin{equation}\label{eq:cond_indep}
 p(r_F|x) = \prod_{j\in F} p(r_j|x),
\end{equation}
where the conditional probability $p(r_F|x) = p(r_F,x)/p(x)$ is defined as usual. Conditional independence guarantees that every elementary fragment $j$ contains new valuable information about $x$ provided that $r_j$ can distinguish between $x$ and $x' (\neq x)$. The last condition means that the conditional probabilities must be different: $p(r_j|x) \neq p(r_j|x')$ for $x\neq x'$.\footnote{If $p(r_j|x)$ is the same for all $x$, then fragment $j$ does not contribute to the MI, i.e., for any $F\supset\{j\}$ we have $I_C(S:R_F) = I_C(S:R_{F\setminus\{j\}})$. While it is possible to tolerate a few of these uninformative fragments $j$, the emergence of a redundancy plateau forbids a large number of them.}

Focusing on the equal time case $s\rightarrow t$, we find that conditional independence is guaranteed by the SBS or branching form. To see this, consider the projector $\Pi_S(x)\otimes\Pi(r_j)\otimes\Pi(r_k)$, where $\Pi_S(x)$ projects on the pointer basis of the system and $\Pi(r_j)$ and $\Pi(r_k)$ on fragment $j$ and $k$, respectively. The joint probability $p(x,r_j,r_k)$ for a state of branching form~(\ref{eq:QD:branching form}) is then
\begin{equation}
 \begin{split}\label{eq:QD-cond-independence}
  p(x,r_j,r_k) &= \lr{\Psi|\Pi_S(x)\otimes\Pi(r_j)\otimes\Pi(r_k)|\Psi}_{SE} \\
  &= p(r_j|x) p(r_k|x) p(x)
 \end{split}
\end{equation}
with $p(r_j|x) = \lr{\varphi_j(x)|\Pi(r_j)|\varphi_j(x)}$. Clearly, $p(x,r_j,r_k)$ satisfies conditional independence. This argument can be generalized to any fragment $F$ and the same conclusion would be reached for a state of SBS. Conversely, conditional independence implies the branching form for record projectors $\Pi(r_j) = |r_j\rl r_j|_j$ that are rank one. To see this, consider the identity 
\begin{equation}
 \begin{split}
  |\Psi\rangle_{SE} &= \sum_{x,r_E} \Pi(x)\otimes\Pi(r_E)|\Psi\rangle_{SE} \\
  &= \sum_{x,r_E} \sqrt{p(x,r_E)} |x\rangle|r_1\rangle\cdots|r_N\rangle.
 \end{split}
\end{equation}
We now use conditional independence to write
\begin{equation}
 |\Psi\rangle_{SE} = \sum_x \sqrt{p(x)} |x\rangle |\varphi_1(x)\rangle \cdots |\varphi_N(x)\rangle
\end{equation}
with $|\varphi_j(x)\rangle \equiv \sum_{r_j} \sqrt{p(r_j|x)}|r_j\rangle$. Since redundancy demands $p(r_j|x)\neq p(r_j|x')$ for $x\neq x'$, we find $\lr{\varphi_j(x)|\varphi_j(x')} < 1$ and confirm the branching form.

It is known that the SBS or branching form imply a redundancy plateau of quantum \emph{and} classical MI, essentially because a SBS has only classical correlations (we will present a proof in Sec.~\ref{sec:extensions}). We will also see in Sec.~\ref{sec:extensions} that conditional independence is sufficient but \emph{not} necessary to imply a redundancy plateau because conditional independence is sufficient but \emph{not} necessary to guarantee that each fragment contains new valuable information about $x$. For the reader interested in the SBS and QD debate, we note that the idea of conditional independence is analogous to the idea of ``strong independence''~\cite{HorodeckiKorbiczHorodeckiPRA2015, KorbiczQuantum2021}.

Thus, past redundancy (in the limit $s\rightarrow t$) is situated in the context of previous works (Sec.~\ref{sec:QD}) as follows. First, past redundancy is a slightly stronger notion that the original formulation of redundancy: it always implies eqn~(\ref{eq:QD:MI}), and it implies that information is locally and classically accessible, but the converse is not true. Second, past redundancy is a weaker notion than the SBS or branching form, as one needs the additional condition of conditional independence to imply eqns~(\ref{eq:QD:SBS}) or~(\ref{eq:QD:branching form}). 

\subsection{Probability sum rule and decoherent histories}\label{sec:classicality}

We turn to second point in the definition of past redundancy: classicality. It is easy to check that the probabilities defined in eqn~(\ref{eq:prob_def}) always satisfy $p(r_F,x)\ge0$ and $\sum_{r_F,x} p(r_F,x) = 1$, but in general the $p(r_F,x)$ lack a property known as the probability sum rule or Kolmogorov consistency condition. Namely, the bare probability to measure $r_F$, unconditioned on any event $(x;s)$, is
\begin{equation}\label{eq:prob_rF}
 p(r_F,\cancel{x}) \equiv \mbox{tr}\{\Pi(r_F)|\Psi_t\rl\Psi_t|\}.
\end{equation}
A problem arises now because the probability sum rule
\begin{equation}\label{eq:prob sum rule}
 p(r_F,\cancel{x}) = \sum_x p(r_F,x)
\end{equation}
is in general \emph{violated} owing to the presence of quantum coherences since $\sum_x\Pi_S(x)|\Psi_s\rl\Psi_s|\Pi_S(x) \neq |\Psi_s\rl\Psi_s|$. 

This is a profound problem since without sum rule it is unclear how to interpret the classical MI $I_C(S:R_F) = H(S) + H(R_F) - H(SR_F)$. If $H(R_F)$ is defined with respect to the marginal $p(r_F) \equiv \sum_x p(r_F,x)$, what is its meaning when $p(r_F)\neq p(r_F,\cancel{x})$? Demanding the validity of eqn~(\ref{eq:prob sum rule}) can thus be seen as a definition of classicality, for further motivation outside the context of redundancy see Refs.~\cite{VanKampenPhys1954, GriffithsJSP1984, AnastopoulosFP2006, SmirneEtAlQST2018, StrasbergDiazPRA2019, MilzEtAlQuantum2020, MilzEtAlPRX2020, EgriEtAlArXiv2026}.

Two quick remarks are in order. First, notice that the probability sum rule with respect to the past event is always obeyed: $p(x,\cancel{r_F}) = \sum_{r_F} p(r_F,x)$. Second, the sum rule is also always obeyed if we consider the equal time limit $s\rightarrow t$. Thus, in our previous discussion in connection to QD (Secs.~\ref{sec:QD_connection} and~\ref{sec:cond_indep}) we did not need to worry about the right definition of classicality. 

Interestingly, there is a slightly stronger notion of classicality: the DH condition (also called the consistent history condition). In our case it is defined as
\begin{equation}\label{eq:DHC}
 |\lr{\psi(r_F,x)|\psi(r_F,x')}| = \delta_{x,x'},
\end{equation}
and it implies the probability sum rule but the converse is not true. Apart from the connection to the existence of formal records of the past~\cite{HartleLecture1992, GellMannHartlePRD1993}, there are also other reasons to prefer the slightly stronger notion of eqn~(\ref{eq:DHC}) over eqn~(\ref{eq:prob sum rule})~\cite{BittnerRosskyJCP1997, DiosiPRL2004, SzankowskiCywinskiQuantum2024}. In context of redundancy, we will discuss differences between the two notions in Secs.~\ref{sec:Haroche} and~\ref{sec:extensions}, but overall we will remain agnostic here about what is the ``right'' notion of classicality. 

In fact, in many physically relevant situations one finds both conditions satisfied to a very good approximation~\cite{DowkerHalliwellPRD1992, GemmerSteinigewegPRE2014, SchmidtkeGemmerPRE2016, AlbrechtBaunachArrasmithPRD2022, StrasbergEtAlPRA2023, StrasbergSP2023, StrasbergReinhardSchindlerPRX2024, WangStrasbergPRL2025, StrasbergEtAlArXiv2026}. Moreover, one naturally expects that environmentally induced decoherence implies the DH condition with respect to the pointer basis---a claim that can be made rigorous if the open system dynamics satisfies a strong notion of quantum Markovianity~\cite{PazZurekPRD1993, StrasbergSP2023}.

However, most important for our context is the insight that the redundancy plateau of eqn~(\ref{eq:classical_plateau}) alone suffices to imply an approximate probability sum rule (and an approximate DH condition under mild assumptions). This will be established in Sec.~\ref{sec:extensions}. 

\subsection{Brief summary}

The notion of an event in the past and the idea that the present has redundant records about this event were defined precisely. In the equal time limit $s\rightarrow t$ past redundancy smoothly connects to previously studied concepts. We argued that for good reasons it is slightly stronger than the definition in terms of quantum MI, but weaker than the condition of a SBS or branching form. Finally, we highlighted the meaning of classicality for redundancy. Importantly, we announced that redundancy does not require classicality, it implies classicality (see Sec.~\ref{sec:extensions}). 

\section{Examples}\label{sec:examples}

We investigate the emergence of past redundancy in three models of increasing complexity. Each model shows eternal past redundancy but only temporary equal-time redundancy. Altogether, this suggests that past redundancy is a stable phenomenon that emerges under wider conditions than QD. This will be also discussed from a broader perspective in the conclusions (Sec.~\ref{sec:conclusions}).

\subsection{CNOT-fan-out circuits with thermalization}

\begin{figure}[t]
	\centering\includegraphics[width=0.26\textwidth,clip=true]{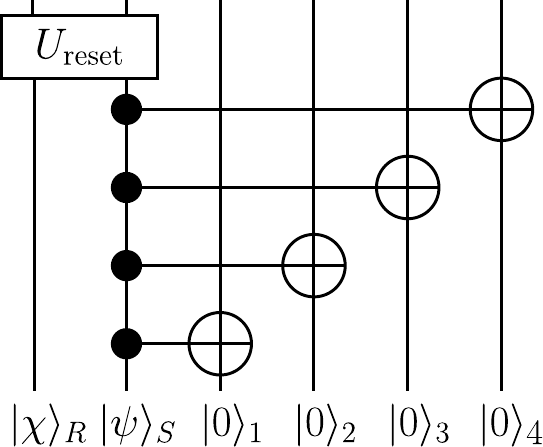}
	\caption{Modified CNOT-fan-out quantum circuit with a final reset operation.}
	\label{fig cnot circuit mod}
\end{figure}

As a first particularly simple but nevertheless instructive example we consider the quantum circuit shown in Fig.~\ref{fig cnot circuit mod}. It is a slight variation of the circuit shown in Fig.~\ref{fig:cnot circuit}, where an additional quantum system $R$ interacts with the system at the end via a unitary gate $U_\text{reset}$, which is programmed to reset the system state to a fixed reference state $\gamma_S$ independent of its prior state. For an initial system state $|\psi_\text{in}\rangle_S = c_0|0\rangle + c_1|1\rangle$, the system-ancilla output state, after tracing out $R$, is
\begin{equation}
 \rho_{SA}^\text{out} = \gamma_S \otimes \left(|c_0|^2 |0000\rl0000|_A + |c_1|^2 |1111\rl1111|_A\right).
\end{equation}
Suddenly, QD ceases to exist and neither eqns~(\ref{eq:QD:MI}),~(\ref{eq:QD:SBS}) or~(\ref{eq:QD:branching form}) is even approximately satisfied. Yet, the ancilla states still reliably encode the state of the system \emph{before} the reset operation $U_\text{reset}$: the past redundancy between the ancilla states and the initial system state remains \emph{unchanged} by the reset operation.

Importantly, while this specific quantum circuit example is simplistic, it suggests a general picture that one can defend as realistic. Any generic environment is characterized by two different types of effects on the system: decoherence and dissipation (or thermal relaxation). The decohering type destroys phase coherence on an immensely fast scale~\cite{JoosZehZPB1985, Zurek1986}, an effect that does not necessarily require any significant energy exchange between the system and the environment. This is captured by the CNOT gates in Fig.~\ref{fig cnot circuit mod}. The thermalizing type instead happens often on a slower time scale and requires exchange of energy, but, importantly, thermalization is as generic as decoherence in our universe. The effect of thermalization, which erases any prior information about the system state and destroys QD, is captured by the $U_\text{reset}$ gate in Fig.~\ref{fig cnot circuit mod}.

In reality, a system is coupled to an environment which causes both decoherence and thermalization to different degrees and at different time scales. In the following, we first treat an example where the same environmental degrees of freedom are responsible for decoherence and thermalization. Afterwards, we treat an example of experimental relevance where both effects are neatly separated.

\subsection{Open cavity in a coherent state}\label{sec:coherent state}

The microscopic model describes a central oscillator ($a,a^\dagger$) coupled to a bath of oscillators ($b_k,b_k^\dagger$) via an interaction within rotating wave approximation:
\begin{equation}\label{eq:Hamiltonian:M2}
 H = \omega_0 a^\dagger a + \sum_{k=1}^N \omega_k b_k^\dagger b_k + \sum_{k=1}^N \gamma_k (a^\dagger b_k + ab_k^\dagger).
\end{equation}
This is a simple and widely used microscopic model to describe an open quantum system such as a lossy optical cavity~\cite{ScullyZubairyBook1997}. Our exposition below uses the notation and parameter choice of Ref.~\cite{OliveiraPaulaDrumondPRA2019}, where this model has been investigated in the context of QD and non-Markovianity but without any relation to past redundancy. 

Owing to the quadratic form of the Hamiltonian~(\ref{eq:Hamiltonian:M2}) one can solve the dynamics exactly. The initial state $|\Psi(0)\rangle = |\alpha(0)\rangle \bigotimes_{k=1}^N|0_k\rangle$ gives rise to the time-evolved state $|\Psi(t)\rangle = |\alpha(t)\rangle \bigotimes_{k=1}^N|\lambda_k(t)\rangle$, where $|\alpha\rangle$ and $|\lambda_k\rangle$ are coherent states (with $|0_k\rangle$ the ground state). The time evolution of the coherent state amplitudes obeys
\begin{align}
 i\dot\alpha(t) &= \omega_0\alpha(t) + \sum_k \gamma_k \lambda_k(t), \\
 i\dot\lambda_k(t) &= \omega_k\lambda_k(t) + \gamma_k \alpha(t).
\end{align}

\begin{figure}[t]
	\centering\includegraphics[width=0.44\textwidth,clip=true]{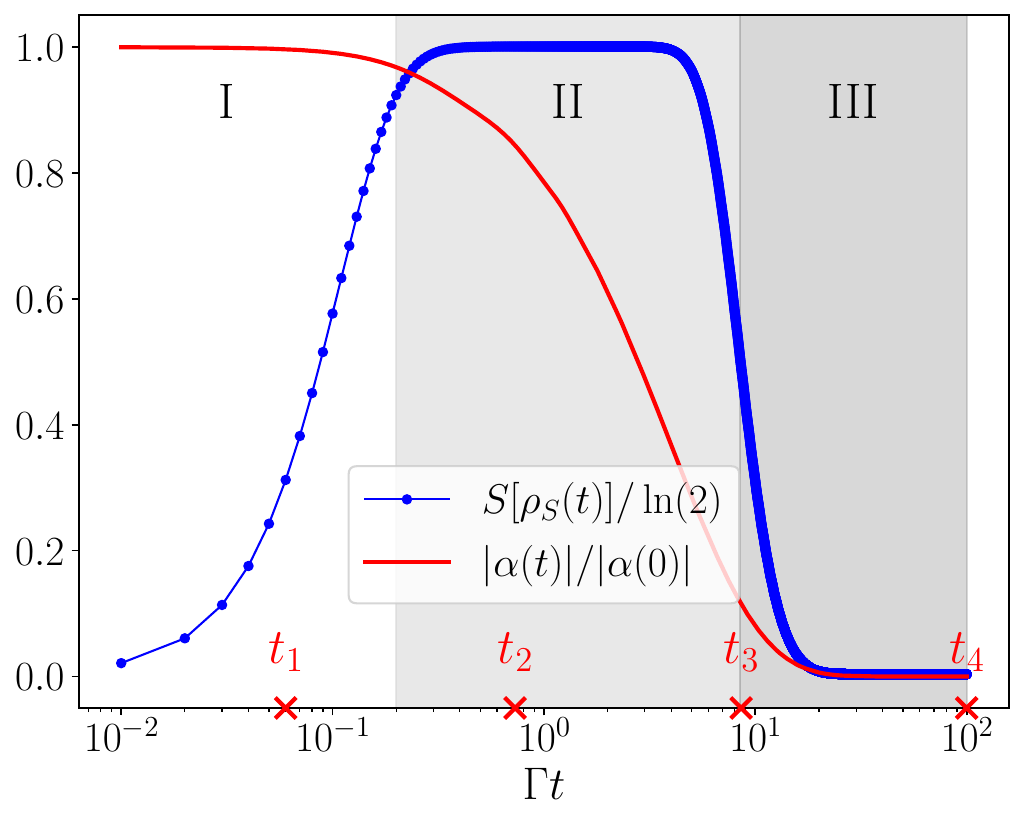}
	\caption{Coherent state amplitude and von Neumann entropy of the open system as a function of dimensionless time (note the logarithmic scale). We set $\omega_0=1$ and consider $N=1000$ bath oscillators with equally spaced frequencies $\omega_k$ in the interval $[0.1,1.9]$ of width $\Delta\omega=1.8$ coupled with equal strength $\gamma_k=\gamma=1/300$ to the system (identical to one parameter choice in Ref.~\cite{OliveiraPaulaDrumondPRA2019}). A good time scale for the plots is set by the rate $\Gamma = 4\pi \gamma^2 N/\Delta\omega$.}
	\label{fig sys ent}
\end{figure}

To investigate redundancy, we pay attention to the following ``Schr\"odinger cat'' initial state 
\begin{equation}
 |\Psi(0)\rangle = \frac{|\alpha(0)\rangle + |\tm\alpha(0)\rangle}{\sqrt{2}\sqrt{1+e^{-2|\alpha(0)|^2}}} \bigotimes_{k=1}^N|0_k\rangle
\end{equation}
with $\alpha(0) = 3$. The extra term in the denominator accounts for proper normalizations since $\lr{\beta|\tm\beta} = e^{-2|\beta|^2}$ for any coherent state $|\beta\rangle$, but note that for $\alpha(0) = 3$ we have almost orthogonality: $\lr{\alpha(0)|\tm\alpha(0)}\approx 1.5 \cdot 10^{-8}$.

To get a feeling for the system dynamics, we first plot the time evolution of the coherent state amplitude $|\alpha(t)|$ and of the system's von Neumann entropy $S_S(t)$ in Fig.~\ref{fig sys ent}, which shows three distinct regimes. For very short times (regime I) we have pure decoherence: a quick loss of coherence gives rise to a quick increase from $S_S(0) = 0$ to $S_S(t)\approx\ln 2$, while $|\alpha(t)|$ remains virtually unaffected. This is followed by a transient regime II in which $S_S(t)\approx\ln2$ stays approximately constant and $|\alpha(t)|$ starts to decay: this is the regime of decoherence and dissipation. Finally, regime III describes the long-time limit. Both the amplitude $|\alpha(t)|$ and von Neumann entropy approach zero and remain there: the system has reached thermal equilibrium.

\begin{figure}[t]
	\centering\includegraphics[width=0.48\textwidth,clip=true]{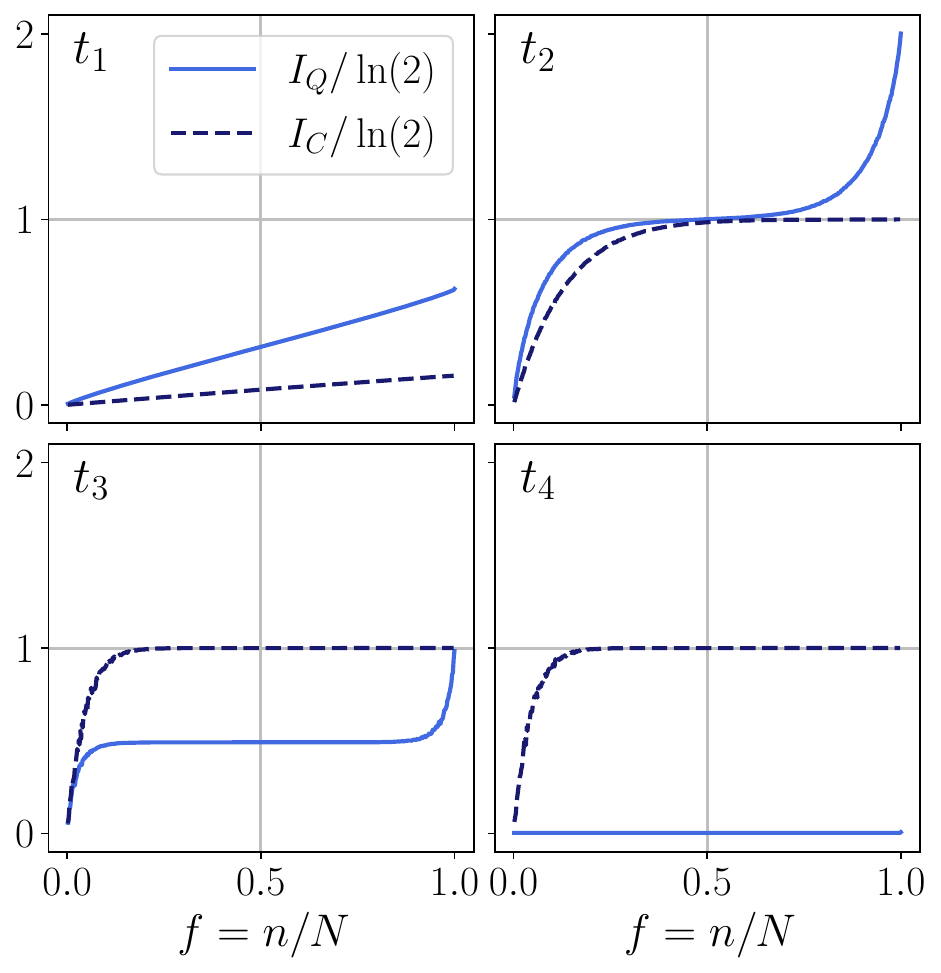}
	\caption{Partial information plots as a function of relative fragment size for four different times $t\in\{t_1,t_2,t_3,t_4\}$ (as indicated in Fig.~\ref{fig sys ent}). Light blue lines: quantum MI in the current state $\rho_{SF}(t)$. Dashed dark blue lines: classical MI between the initial system state and the final record states. }
	\label{fig PIP}
\end{figure}

It follows from the linearity of the Schr\"odinger equation that the time-evolved state is
\begin{equation}\label{eq:E3:branching form}
 |\Psi(t)\rangle \sim |\alpha(t)\rangle \bigotimes_{k=1}^N|\lambda_k\rangle + |\tm\alpha(t)\rangle \bigotimes_{k=1}^N|\tm\lambda_k\rangle.
\end{equation}
Remarkably, this state appears to be of branching form for all times $t\neq0$. However, the stored information in the records, after a quick initial rise, monotonically decreases to zero owing to the decay and non-orthogonality of $|\text{$\pm$}\alpha(t)\rangle$. This is confirmed by the characteristic partial information plots of QD in Fig.~\ref{fig PIP}, displayed for four different times $\{t_1,t_2,t_3,t_4\}$ marked by red crosses in Fig.~\ref{fig sys ent}. These plots are generated by sampling 100 fragments $F$ of a given size $n = fN \in \{1,\dots,N\}$ (with $f$ the relative fragment size) and by computing the quantum MI $I_Q(S:F)$. The empirical average over all samples $F$ then gives an average quantum MI $\lr{I_Q(S:F)}$, which is plotted in Fig.~\ref{fig PIP}. We can clearly see that only in regime II the fragments of the environment allow to discriminate well different system states.

In contrast, in regime III the records in the environment reveal little about the \emph{current} state of the system, but they continue to reliably encode the \emph{initial} state of the system. This is captured by the concept of past redundancy. To see this, choose the event we like to distinguish as $x\in\{\pm\}$ with $\Pi(+) = |\alpha(0)\rl\alpha(0)|$ and $\Pi(-) = |\tm\alpha(0)\rl\tm\alpha(0)|$ at time $s=0$.\footnote{Any time $s>0$ would also work as long as we stay away from regime III. Moreover, we neglect the tiny overlap $\Pi(+)\Pi(-)\approx 1.5\cdot 10^{-8} |\alpha(0)\rl\tm\alpha(0)|$ for simplicity.} The record projectors $\Pi(r_F)$ on fragment $F$ can be chosen in various ways and we consider the \emph{optimal} quantum state discrimination strategy detailed in Appendix~\ref{sec:app:oscillators}. Analogous to the quantum case we then consider partial information plots for the empirical sample average of the classical MI $\lr{I_C(X:R_F)}$.

The difference between the quantum and classical partial information plots in Fig.~\ref{fig PIP} is striking and revealing. For short times $t_1$ and $t_2$ (in regime I and II, respectively) the quantum MI is larger than the classical MI, which illustrates the fact that quantum MI contains locally inaccessible information. At late times ($t_3$ and $t_4$ in regime III), however, it starts to matter whether one considers the present-present correlations as in $I_Q(S:F)$ or the past-present correlations as in $I_C(X:R_F)$---not only is the classical MI larger than the quantum MI, but in contrast to the quantum MI it is also \emph{not} decaying with time. This captures the essence of our approach for this models: environmental fragments preserve the information about the past and not about the present.

A word of caution is, however, necessary. For the present choice of the initial coherent state amplitude, $\alpha(0)=3$, coherences between different histories start to matter if we consider fragments whose relative size $f\gtrsim0.5$ comprises roughly half of the environment or more (for $f\approx0.5$ the probability sum rule is violated by approximately 1\%). This is computed in Appendix~\ref{sec:app:oscillators}. Clearly, having access to half of the environment (or more) is rather unreasonable.

Finally, we note that the Hamiltonian of eqn~(\ref{eq:Hamiltonian:M2}) does not generate a unitary time evolution of the form~(\ref{eq:U CNOT-fan-out}) and it also does not reduce to the canonical form of QD Hamiltonians given in Ref.~\cite{DoucetDeffnerPRX2024}. The reason is that eqn~(\ref{eq:Hamiltonian:M2}) has no well-defined time-independent pointer basis: the set of coherent states $\{|\alpha\rangle\}$ has no preferred direction in Hilbert space and it is not orthogonal. Nevertheless, to a good approximation coherent states can share many properties of a pointer basis~\cite{ZurekHabibPazPRL1993}. It is an advantage of the present approach that such debates do not matter for the emergence of past redundancy.

\subsection{Quantum non-demolition measurement of Fock states}\label{sec:Haroche}

\begin{figure}[tb]
	\centering\includegraphics[width=0.49\textwidth,clip=true]{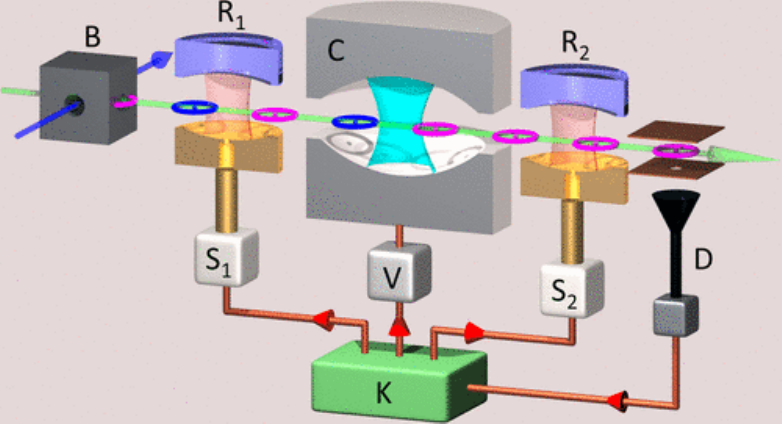}
	\caption{Micromaser setup from Ref.~\cite{ZhouEtAlPRL2012}: A microwave cavity {\texttt C} in the middle interacts with a stream of atoms prepared and read out using an oven {\texttt B}, Ramsey cavities {\texttt R}$_1$ and {\texttt R}$_2$, and an ionization detector {\texttt D}. In the original experiment the information from the pink atoms was used to feedback control the cavity state with the blue atoms~\cite{ZhouEtAlPRL2012}, but here we do \emph{not} consider any feedback control.}
	\label{fig Haroche}
\end{figure}

We consider a micromaser setup as the one shown in Fig.~\ref{fig Haroche}, which was used in the lab of Haroche to implement quantum non-demolition measurements of the photon number of a microwave cavity~\cite{GuerlinEtAlNature2007, ZhouEtAlPRL2012, HarocheRMP2013}. The relevant physics can be summarized as follows.

First, the microwave cavity is an open system characterized by photon losses to an uncontrollable environment.\footnote{We remark that the master equation governing the cavity evolution [eqn~(\ref{eq:ME}) in Appendix~\ref{sec:app:Haroche}] can be microscopically derived from the previous model in Sec.~\ref{sec:coherent state}~\cite{ScullyZubairyBook1997}.} In our numerical simulation below, we consider an initial cavity state written in the Fock basis as
\begin{equation}\label{eq:Haroche:psi_initial}
 |\psi(0)\rangle_{S} = \frac{|0\rangle_{S} + |2\rangle_{S}}{\sqrt{2}},
\end{equation}
i.e., a superposition of zero and two photons. The lossy cavity causes an exponential decay of the photon number $\lr{n}\sim e^{-t/\tau_c}$ (with $\tau_c$ the cavity lifetime) to a state with zero photons.\footnote{For simplicity, we assume a zero temperature environment, which well approximates the experimental conditions that lead to a thermal occupation of $\lr{n}_\text{th}\approx0.05$~\cite{GuerlinEtAlNature2007, ZhouEtAlPRL2012}.} Clearly, the loss of two photons will not generate any significant amount of redundancy in the electromagnetic field modes coupled to the cavity.

However, during the decay of the cavity, the experiment uses an additional stream of atoms, prepared in a state $|\varphi_0\rangle$, that interact sequentially with the cavity (pink and blue rings in Fig.~\ref{fig Haroche}, but the different color does not matter here). Moreover, the interactions happen suddenly on a timescale approximately 2000 times smaller than the cavity lifetime. Hence, the atom-cavity dynamics can be well approximated by an instantaneous unitary map $V$ (where the subscript $j$ labels the $j$'th atom):
\begin{equation}\label{eq:Haroche:interaction}
 V|n\rangle_{S}|\varphi_0\rangle_j = |n\rangle_{S}|\varphi(n)\rangle_j.
\end{equation}
Importantly, the experiment is designed such that the number of cavity photons $n$ does not change during the interaction (known as the dispersive coupling regime in quantum optics~\cite{ScullyZubairyBook1997}). Note that this is exactly the type of unitary map from eqn~(\ref{eq:U CNOT-fan-out}).

Independent of any further details we can now see a competition arising. If we could switch off the photon losses to the environment, the cavity photon number would be indefinitely (and hence redundantly) encoded in the state of the atoms. In the opposite case of very strong dissipation (i.e., for a very lousy cavity), all the photons might be lost before a significant amount of atoms had the chance to interact with the cavity.

\begin{figure}[t]
	\centering\includegraphics[width=0.45\textwidth,clip=true]{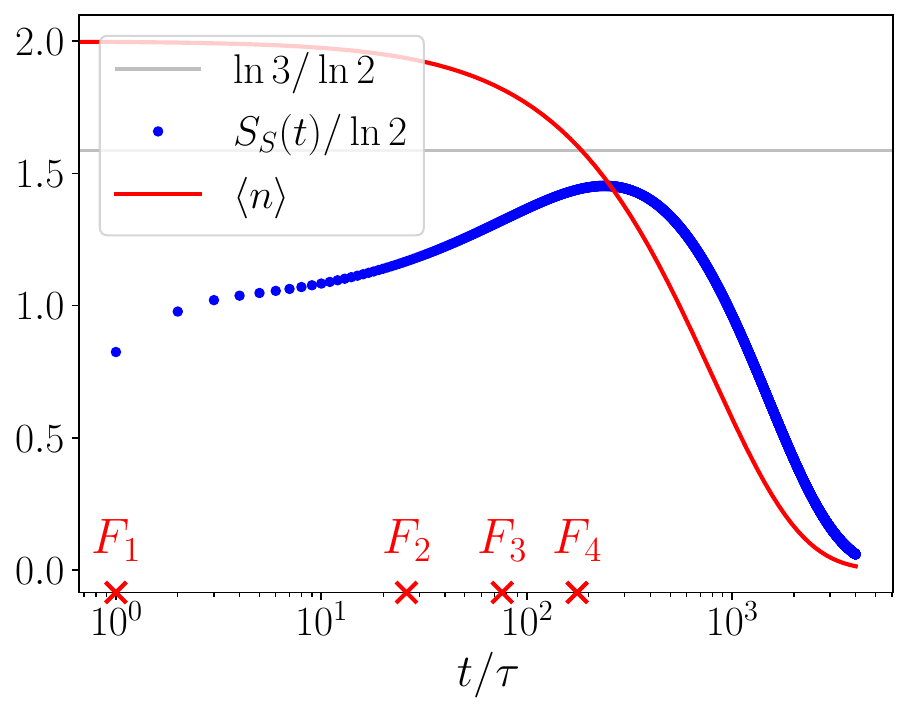}
	\caption{As a function of time $t/\tau$ in logarithmic scale: expected photon number in the cavity $\lr{n}(t)$ (red solid line) and system entropy $S_S(t)$ (blue discs). The red crosses on the $x$-axis indicate at which time different atom fragments start to interact with the cavity, see the main text for details.}
	\label{fig Haroche S n}
\end{figure}

Figure~\ref{fig Haroche S n} gives insights into this competition by plotting two quantities for the experimentally realistic case where the period $\tau$ in between two consecutive atoms is $\tau=\tau_c/800$~\cite{ZhouEtAlPRL2012}.\footnote{For the current pure measurement (no feedback) setup there are essentially four free parameters to choose: the dimensionless ratio $\tau_c/\tau$ (set to 800), the dimensionless temperature to cavity frequency ratio $k_BT/\hbar\omega_0$ (set to $T=0$), the initial cavity state [see eqn~(\ref{eq:Haroche:psi_initial})], and the atom output states $|\varphi(n)\rangle_j$ in eqn~(\ref{eq:Haroche:interaction}), which we specify in Appendix~\ref{sec:app:Haroche}.} To better resolve the time scales involved in the problem, we use a logarithmic scale for time.

First, we display the expected photon number $\lr{n}(t)$ (red solid line) as a function of time for an initial state $|\psi(0)\rangle_C = |2\rangle_C$ with two photons. It shows the familiar exponential decay, here obscured by the logarithmic time scale. What matters is that we consider times long enough to see an almost fully decayed cavity state. Specifically, we consider 4000 cycles of duration $\tau$, i.e., 4000 atoms interacting sequentially with the cavity. Since $\tau_c=800\tau$, we have $t\in[0,5\tau_c]$, which implies a final photon number of $\lr{n}(5\tau_c) = e^{-5} \approx 0.0067$, i.e., the cavity finishes basically in the ground state.

Second, we display the system entropy $S_S(t)$ (blue discs) as a function of time for the initial state in eqn~(\ref{eq:Haroche:psi_initial}). It features a sharp initial rise (recall that $S_S(0) = 0$) owing to the entanglement with the first atoms. Moreover, its maximum is close to $\ln(3)$ owing to the fact that the relevant Hilbert space of the cavity, $\mbox{span}\{|0\rangle,|1\rangle,|2\rangle\}$, is three-dimensional. Finally, for long times it decays to zero owing to the dissipation of the cavity.

\begin{figure}[t]
	\centering\includegraphics[width=0.45\textwidth,clip=true]{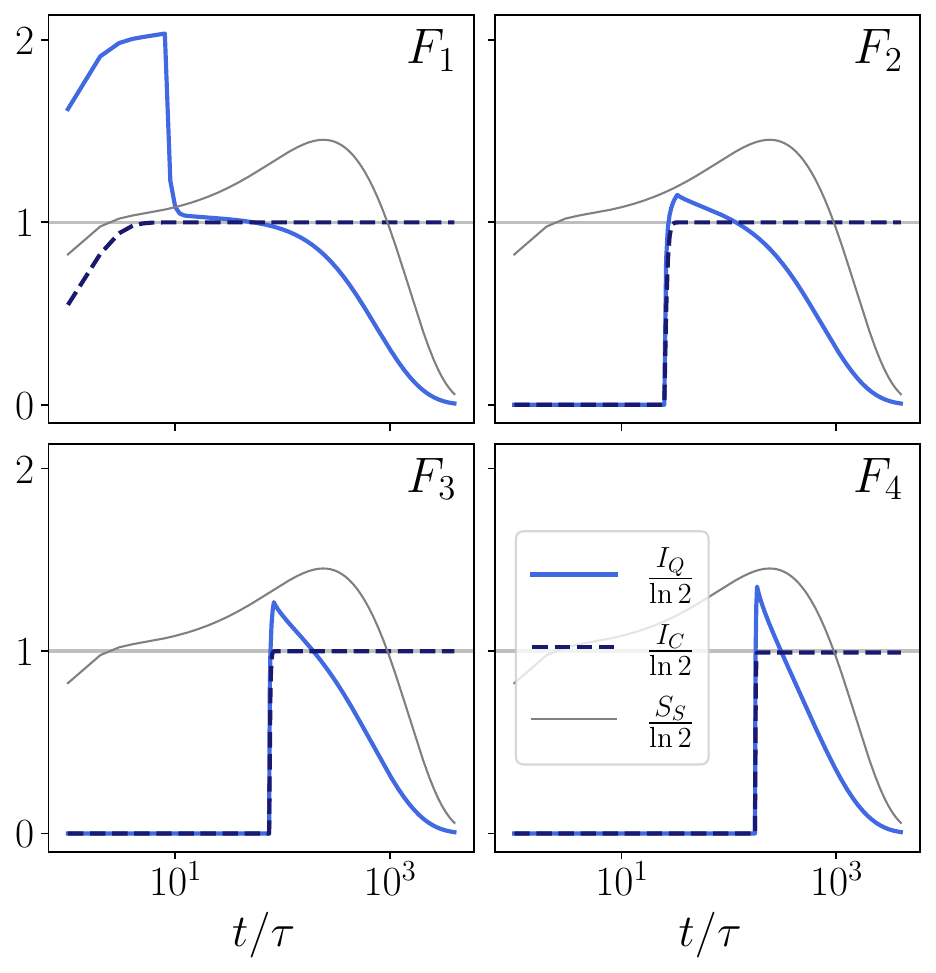}
	\caption{As a function of time $t/\tau$ in logarithmic scale: quantum MI $I_Q(S:F_i)$ (light blue lines) for different fragments $F_1 = \{1,\dots,8\}$, $F_2 = \{26,\dots,33\}$, $F_3 = \{76,\dots,83\}$ and $F_4 = \{176,\dots,183\}$, and the corresponding classical MI $I_C(X:R)$ for the same fragments (dashed dark blue lines). }
	\label{fig Haroche MI}
\end{figure}

To investigate QD and past redundancy we turn to Fig.~\ref{fig Haroche MI}, where we study the time evolution of the cavity-fragment correlation for four different fragments $F\in\{F_1,F_2,F_3,F_4\}$ of eight atoms each. Specifically, these fragments are $F_1 = \{1,\dots,8\}$, $F_2 = \{26,\dots,33\}$, $F_3 = \{76,\dots,83\}$ and $F_4 = \{176,\dots,183\}$. The time at which the first atom of each fragment interacts with the cavity is marked by a red cross in Fig.~\ref{fig Haroche S n}.

We begin with the discussion of quantum MI $I_Q(S:F_j)$ (light blue solid lines). The first fragment is somewhat peculiar as $I_Q(S:F_1)$ rises above the classically allowed value $S_S(t)$ (thin grey line) for short times: owing to the weak dissipation, the dynamics is almost unitary for short times such that $I_Q(S:F_1)$ contains a significant amount of quantum entanglement. However, for times $t/\tau>8$ there is a quick decay to $I_Q(S:F_1)\approx\ln2$ owing to the destruction of quantum coherences, caused primarily by the subsequent atoms $j\in\{9,10,\dots\}$ and not the cavity dynamics which still only weakly influences the state [compare with $\lr{n}(t)$]. After that initial phase, there is a rather large transient time window for which $I_Q(S:F_1)\approx\ln2$ remains approximately constant, before it eventually decays to zero owing to the photon losses. Instead, $I_Q(S:F_i)$ for the other three fragments $i\in\{2,3,4\}$ does not rise above the classically allowed value $S_S(t)$. However, it comes close to the classical limit, in agreement with eqn~(\ref{eq:QD:MI}), before quickly decaying to zero. The overall conclusion is therefore: the atoms that interact with the cavity during a time window comparable to the cavity life time show a moderate signature of QD, but eventually any signature of QD is \emph{lost}.

However, even when any correlation between the atoms and the \emph{present} cavity state is lost, the atoms still faithfully encode the \emph{past} state of the cavity. To confirm this idea, we computed the classical MI $I_C(X:R_{F_i})$. Here, $X = \{(x;0)\}$ denotes the event at time $s=0$ which encodes the number of photons $x\in\{0,2\}$ in the cavity (with corresponding projectors $|0\rl0|_S$ and $|2\rl2|_S$) and $R_{F_i} = \{(r_{F_i};t)\}$ denotes the record random variable for the same fragments $F_i$ as above. The probability distribution for $R_{F_i}$ is generated with respect to records encoding the ground or excited state of every atom in fragment $F_i$, as measured in the experiment. We defer the details to Appendix~\ref{sec:app:Haroche}, where we also show that the probability sum rule~(\ref{eq:prob sum rule}) is obeyed for all fragments at all times, and here focus only on the resulting phenomenology, which shows two important characteristics in Fig.~\ref{fig Haroche MI}.

First, the rise of classical MI is in general slower than the rise of quantum MI because the latter captures more correlations than the former. Unfortunately, this observation can not be elevated to a theorem since $I_C(X:R_{F_i}) \le I_Q(S:F_i)$ only holds if $X$ and $R_{F_i}$ are measured at the \emph{same} time. However, as a rule of thumb it seems justified to assume that $I_C(X:R_{F_i}) \le I_Q(S:F_i)$ for short times. Second, and more importantly, the past-present correlations encoded in $I_C(X:R_{F_i})$ do not decay to zero: they remain constant after the last atom of the fragment has interacted with the cavity. This is a crucial feature of past redundancy: the information stored in the atoms does not become worthless---as a direct application of the various definitions in Sec.~\ref{sec:QD} would suggest---but instead they encode forever the history of the system (until, of course, some other here not considered mechanism changes the state of the atoms).

\section{Extensions and General Results}\label{sec:extensions}

We continue to further develop the theory of past redundancy. We provide the long promised proof that a redundancy plateau in classical MI implies classicality in Sec.~\ref{sec:dependency}. In addition, we also demonstrate that conditional independence (see Sec.~\ref{sec:cond_indep}) is not needed for past redundancy. Section~\ref{sec:multiple events} extends the formalism to multiple past events. 

The \emph{Bhattacharyya coefficient} (BC) $B(q,p)$ between two probability distributions $\{p(a)\}$ and $\{q(a)\}$ plays an important role below~\cite{Bhattacharyya1943}:
\begin{equation}
 B(p|q) \equiv \sum_a \sqrt{p(a)q(a)}.
\end{equation}
We have $B(p|q) = B(q|p) \in[0,1]$ with $B(q|p)=0$ if and only if $q$ and $p$ have orthogonal support and $B(q|p)=1$ if and only if $p=q$. The BC is a sort of distance measure between two probability distributions, but it does not satisfy the triangle inequality. We remark that the quantum analogue of the BC is the fidelity $F(\rho|\sigma) \equiv \mbox{tr}\{(\sqrt{\rho}\sigma\sqrt{\rho})^{1/2}\}$~\cite{NielsenChuangBook2000, WildeBook2019}. Since we only need the classical BC in the following, we keep the letter $F$ to exclusively denote fragments of the environment.

\subsection{Redundancy, classicality and correlated records}\label{sec:dependency}

\subsubsection{Setup and redundancy definition}

The goal is to define past redundancy between an event random variable $X$ at time $s$ and record random variables $R_F$ at time $t$ for fragments $F$ with respect to local projectors compatible with the tensor product decomposition of the environment $\C H_E = \bigotimes_j \C H_j$. This is identical to the setting of Sec.~\ref{sec:basics}, but we now add one important generalization. So far, to keep the discussion closely aligned with the QD literature, $X$ was a random variable associated to events defined by projectors $\{\Pi_S(x)\}_{x=1}^m$ on the open system (and to make the connection with QD we assumed it to coincide with the pointer basis). However, in reality we can use an \emph{arbitrary} complete set of orthogonal projectors $\{\Pi(x)\}_{x=1}^m$, no distinguished open system $S$ is ever needed. In fact, systems can and often do disappear throughout evolution~\cite{GellMannHartlePRD1993, RiedelPC2026}, so it is important to liberate us from this restriction. This is difficult to achieve within the traditional setting of QD as the notion of a reduced density matrix $\rho_S(t)$ for the system is indispensable to even formulate the criteria in Sec.~\ref{sec:QD}. Within an approach rooted in the DH formalism we can realize, however, that this was never needed to define an event random variable.

We now define past redundancy only by condition~1. from Sec.~\ref{sec:first_def}, repeated here for convenience:

\bb{(1)} The classical MI $I_C(X:R_F)$ exhibits a classical \emph{redundancy plateau}, i.e., $I_C(X:R_F) \lesssim H(X)$ for all fragments $F$ that are sufficiently large but still much smaller than the total environment: $|F|\ll N$.

We remark that for technical reasons we assume below that $p(x) > 0$ for all events $x$ or, put differently, we have to exclude those $x$ for which $p(x)=0$. We can not make any statements about impossible events that never happened (owing, e.g., to a specific initial state $|\Psi_0\rangle$).

\subsubsection{Sandwich theorem}

We now state the first main result of this section, which links the redundancy plateau to the BC 
\begin{equation}
 B_F(x,x') \equiv \sum_{r_F} \sqrt{p(r_F|x)p(r_F|x')} 
\end{equation}
of the conditional record distributions $p(r_F|x)$ for fragment $F$ for different $x\neq x'$. Since this result is so central, we promote it to a Theorem: \\

\begin{thm*}[Sandwich]
 The conditional entropy $H(X|R_F)$ is sandwiched by the BC. A lower bound is given by 
 \begin{equation}\label{eq:bound1}
  H(X|R_F) \ge qH_2\left[\frac{1-\sqrt{1-4p_1(x)p_1(x')B_F(x|x')}}{2}\right],
 \end{equation}
 where $H_2(r) = -r\ln(r)-(1-r)\ln(1-r)$ is the binary Shannon entropy, and we defined $q \equiv p(x) + p(x')$, $p_1(x) \equiv p(x)/q$ and $p_1(x') \equiv p(x')/q$. An upper bound is given by
 \begin{equation}\label{eq:bound2}
  H(X|R_F) \le H_2(P_\text{\normalfont error}) + P_\text{\normalfont error}\ln(m-1),
 \end{equation}
 where $P_\text{\normalfont error} \le \sum_{x\neq x'}\sqrt{p(x)p(x')} B_F(x,x')$ is an error probability.
\end{thm*}

The proof of this theorem is detailed in Appendix~\ref{sec:app:proofs}. Essentially, it is a generalization of a result given in Ref.~\cite{FuchsVanDeGraafIEEE1999} for binary random variables $X$. The lower bound~(\ref{eq:bound1}) follows from Ref.~\cite{FuchsVanDeGraafIEEE1999} with some modifications. The upper bound~(\ref{eq:bound2}) uses results from Refs.~\cite{Bhattacharyya1943, KailathIEEE1967, CoverThomasBook1991}. It is based on a bound of the discrimination error $P_\text{error}$ in a Bayesian hypothesis testing scenario in terms of the BC. An application of Fano's inequality then gives eqn~(\ref{eq:bound2}).

To highlight the importance of the sandwich theorem, recall that the gap $H(X) - I_C(X:R_F) = H(X|R_F)$ is controlled by $H(X|R_F)$. Hence, when the gap closes we must have $B_F(x,x') \rightarrow 0$ for all $x\neq x'$ as per eqn~(\ref{eq:bound1}). Similarly, when $B_F(x,x') \rightarrow 0$ for all $x\neq x'$ it follows that $I_C(X:R_F) \rightarrow H(X)$ as per eqn~(\ref{eq:bound2}).

\subsubsection{Strongly decaying BC and (de)correlated records}

The sandwich theorem motivates to introduce the notion of a \emph{strongly decaying family of BCs} $B_F(x|x')$. It is defined for an ordered set of fragments $\C F = \{F_0,F_1,\dots,F_K\}$, which satisfies $F_k\subsetneq F_{k+1}$ for all $k\in\{1,\dots,K-1\}$ and where we include the empty set $F_0 \equiv\emptyset$ and the entire environment $F_K = E$ in $\C F$. We then demand that for any \emph{fixed} $F\in \C F$ it holds that $B_{\overline F}(x|x')/B_F(x|x')$ decays quickly to zero with increasing extended fragment size $|\overline F|$, where $F\subset\overline F\in\C F$. The BC for the empty fragment $F_0$ is set to $B_\emptyset(x,x') \equiv 1$.

To understand this definition, we first notice that $B_{\overline F}(x|x')/B_F(x|x')\le 1$ is always true. This follows from the monotonicity of quantum fidelity~\cite{NielsenChuangBook2000, WildeBook2019}. Next, let us return to the case of conditionally independent records from Sec.~\ref{sec:cond_indep}. From the definition~(\ref{eq:cond_indep}) it immediately follows that
\begin{equation}\label{eq:BC-exp-decay}
 B_F(x,x') = \prod_{j\in F} B_j(x,x') \le C e^{-cn}
\end{equation}
for suitable constants $c$ and $C$ and with $n=|F|$ the size of fragment $F$. Furthermore, the BC also factorizes: $B_{\overline F}(x,x') = B_F(x,x') B_{\overline F\setminus F}(x,x')$. By virtue of the upper bound~(\ref{eq:bound2}) it follows that conditionally independent records imply a convergence of the redundancy plateau $I_C(X:R_F) \rightarrow H(X)$ with a rate proportional to $ne^{-cn}$.

The important point to realize is that milder conditions than conditional independence suffice to imply $I_C(X:R_F) \rightarrow H(X)$. For instance, the BC need not factorize, but an exponential decay might be still possible for fragments $F$ with weak, finite range correlations. Moreover, even a much slower decay (e.g., a power law decay of the form $\sim n^{-a}$) is permissible too if $N$ is large enough. Clearly, this goes well beyond the record statistics predicted by a SBS or branching form without jeopardizing a redundancy plateau for classical MI.

Finally, we remark that the sandwich theorem also allows the converse conclusion. If $I_C(X:R_F) \rightarrow H(X)$, then clearly $B_F(x,x') \rightarrow 0$ too for increasing $|F|$ by virtue of the lower bound~(\ref{eq:bound1}). Moreover, unless $I_C(X:R_F) = H(X)$, which happens in the rather unrealistic case that the record of fragment $F$ is perfect and determines $X$ with certainty, it will also hold that $B_{\overline F}(x|x')/B_F(x|x') \rightarrow 0$ for increasing $|\overline F|$ and fixed $|F|$.

\subsubsection{Proof of the probability sum rule}

The reason to introduce the BC becomes clear in the proof of classicality. Let us first consider the probability sum rule in eqn~(\ref{eq:prob sum rule}). A natural way to quantify violations of it uses the Kolmogorov distance:\footnote{The Kolmogorov distance is a true distance or metric (not like the BC), it has many desirable properties in hypothesis testing, and it is the classical analogue of the trace distance in quantum information theory~\cite{NielsenChuangBook2000, WildeBook2019}.}
\begin{equation}
 \Delta_F \equiv \frac{1}{2}\sum_{r_F} \left|p(r_F,\cancel{x}) - \sum_x p(r_F,x)\right| \in [0,1].
\end{equation}
Now, from the definition of the involved probabilities, eqns~(\ref{eq:prob_def}) and~(\ref{eq:prob_rF}), we find
\begin{equation}
 \Delta_F = \frac{1}{2}\sum_{r_F} \left|\sum_{x\neq x'} \sqrt{p(x)p(x')} \lr{\psi_t(x_s)|\Pi(r_F)|\psi_t(x'_s)}\right|.
\end{equation}
Inserting the identity $I_{\C H} = \sum_{r_{\overline F\setminus F}} \Pi(r_{\overline F\setminus F})$, where $r_{\overline F\setminus F}$ is defined such that $r_{\overline F} = (r_{\overline F\setminus F},r_F)$, yields
\begin{equation}
 \begin{split}\label{eq:help_confusion}
  &\Delta_F = \\
  &\frac{1}{2}\sum_{r_F} \left|\sum_{x\neq x'} \sqrt{p(x)p(x')} \sum_{r_{\overline F\setminus F}}\lr{\psi_t(x_s)|\Pi(r_{\overline F})|\psi_t(x'_s)}\right|.
 \end{split}
\end{equation}
Since $\Pi(r_{\overline F})|\psi_t(x_s)\rangle = \sqrt{p(r_{\overline F}|x)}|\psi(r_{\overline F},x_s)\rangle$, we find after applying the triangle and Cauchy-Schwarz inequality the central result that for any $\overline F\supset F$
\begin{equation}
 \begin{split}
  \Delta_F &\le \frac{1}{2} \sum_{x\neq x'} \sqrt{p(x)p(x')} B_{\overline F}(x|x') \\
  &\le m \max_{x,x'} B_{\overline F}(x|x').
 \end{split}
\end{equation}
\emph{Et voil\`a}, the BC controls violations of the probability sum rule, as claimed. Four remarks are in order.

First, it is important to have a bound in terms of the BC of the extended fragment $\overline F$ because $B_j(x|x')$ of a single elementary fragment $j$ might be very small while we still like to ensure that $p(r_j,x)$ can be interpreted classically. Second, the above bound seems to be the only one that features solely the BC and does not rely on any properties beyond the information contained in the record statistics $p(r_F,x)$. Third, this implies that in reality one can expect the bound to be loose, as the application of the triangle and Cauchy-Schwarz inequality assume a worst case scenario.

The last and most important remark is that this derivation is \emph{not} circular because one could object that we used classical probability theory to derive the sandwich theorem to show that the redundancy plateau implies vanishing BCs, which are then claimed to justify classical probabilities in retrospective. The correct viewpoint is the following. First, while the physical meaning of $p(r_F,x)$, its marginals $p(x)$ and $p(r_F)$, and its conditionals $p(r_F|x)$ and $p(x|r_F)$ is \emph{a priori} unclear, they are all mathematically well-defined probabilities. The sandwich theorem implies relations between them by pure mathematics, independent of any interpretation. Now, notice the important fact that the norm of the vector $\Pi(r_F)|\psi_t(x_s)\rangle$, which appears in the derivation of the sum rule after applying the Cauchy-Schwarz inequality to eqn~(\ref{eq:help_confusion}) (and with $F$ replaced by $\overline F$), \emph{coincides} with the definition $p(r_F|x) = p(r_F,x)/p(x)$ used in the sandwich theorem. This coincidence is caused by the fact that the sum rule is always obeyed with respect to the past event, $p(x,\cancel{r_F}) = \sum_{r_F} p(r_F,x)$, such that there is no ambiguity in $p(x)$ and $p(r_F|x)$.

\subsubsection{Decoherent histories and a (counter)example}

We already remarked in Sec.~\ref{sec:classicality} that the probability sum rule is a slightly weaker notion of classicality than the decoherence of (individual) histories [see eqn~(\ref{eq:DHC})]. Within the context of past redundancy, let us illustrate this with a simple counterexample, which obeys the probability sum rule (approximately), but does not show redundancy and whose histories do not decohere.

\begin{figure}[t]
	\centering\includegraphics[width=0.45\textwidth,clip=true]{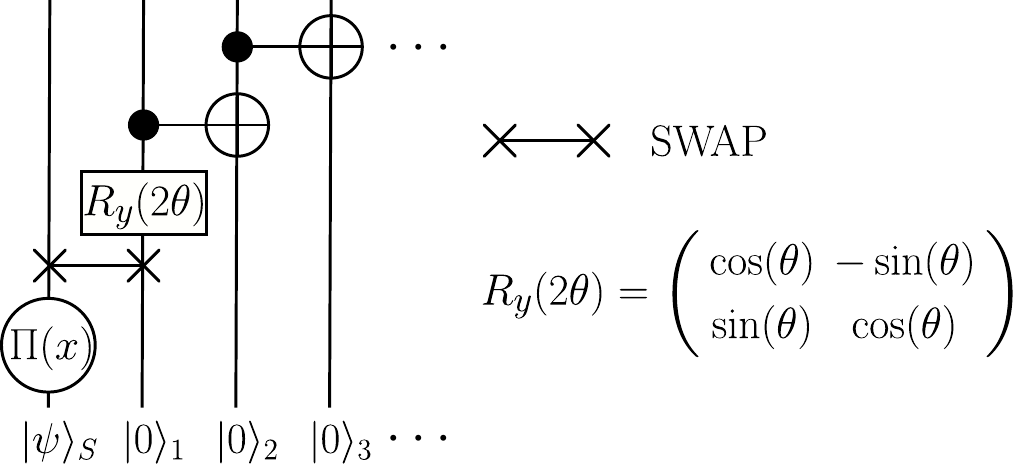}
	\caption{Circuit implementing the counterexample for $|\psi\rangle_S = (|0\rangle_S + |1\rangle_S)/\sqrt{2}$ and $\sin^2(\theta)=\lambda$.}
	\label{fig:circuit counterexample}
\end{figure}

Starting with a system in the state $|\psi(0)\rangle_S = (|0\rangle_S + |1\rangle_S)/\sqrt{2}$ and event projectors $\Pi(x) = |x\rl x|_S$ for $x\in\{0,1\}$ at the initial time, imagine the final branches of the global wave function $|\Psi(t)\rangle = \sum_x |\psi(x)\rangle/\sqrt{2}$ for an environment composed out of $N$ qubits to be
\begin{align}
 |\psi(0)\rangle &= |0\rangle_S\otimes \left[\sqrt{1-\lambda}|0\rangle^{\otimes N} + \sqrt{\lambda}|1\rangle^{\otimes N}\right], \\
 |\psi(1)\rangle &= |0\rangle_S\otimes \left[-\sqrt{\lambda}|0\rangle^{\otimes N} + \sqrt{1-\lambda}|1\rangle^{\otimes N}\right].
\end{align}
A quantum circuit implementation is shown in Fig.~\ref{fig:circuit counterexample}. Now, for $\lambda\gtrsim0$ each qubit in the environment is almost perfectly correlated with the system with respect to the computational basis. The elementary fragment BC is $B_j(x|x') = 2\sqrt{\lambda(1-\lambda)}\gtrsim0$ for all $j$. However, there are \emph{not} $N$ redundant records because two environmental qubits are perfectly correlated in the computational basis, $p(r_j|r_i) = \delta_{r_j,r_i}$, such that $B_{\overline F}(x|x') = B_F(x|x')$ for every $F\subset \overline F$. Thus, even though each environmental qubit knows about the system, there is no redundancy.  The violation of the probability sum rule is now exactly equal to (and not only upper bounded by) the BC
\begin{equation}
 \Delta_F = \frac{1}{2}B_F(x|x') = \sqrt{\lambda(1-\lambda)}.
\end{equation}
This is due to the case that something unusual happens to the histories. Consider, e.g., the two history states for $x\in\{0,1\}$ conditioned on the first environmental qubit being in state $|0\rangle_1$:
\begin{align}
 |\psi(r_1=0,x=0)\rangle &= |0\rangle_S\otimes|0\rangle^{\otimes N}, \\
 |\psi(r_1=0,x=1)\rangle &= |0\rangle_S\otimes|0\rangle^{\otimes N}.
\end{align}
They are equal, and there is no decoherence of histories. Thus, even though the second branch $|\psi(r_1=0,x=1)\rangle$ has a much lower weight $\lambda\gtrsim0$ compared to the first branch, observers finding record $|0\rangle_1$ could not be sure that they are on the branch belonging to event $x=0$.

This example teaches us that a redundancy plateau or a strongly decaying family of BCs does not imply decoherence of all individual histories. This can be seen more generally. First, note that
\begin{equation}\label{eq:BC-relation}
 B_{\overline F}(x|x') = B_F(x|x') \mathbb{E}\left[B_{\overline F\setminus F}(r_F,x|r_F,x')\right],
\end{equation}
where $\mathbb{E}[\dots]$ denotes an average over $r_F$ with respect to the probability distribution $w(r_F) = \sqrt{p(r_F|x)p(r_F|x')}/B_F(x|x')$. Thus, a strongly decaying family of BCs implies that the
\begin{equation}
 \begin{split}\label{eq:cond-BC}
  &B_{\overline F\setminus F}(r_F,x|r_F,x') = \\
  &~~~ \sum_{r_{\overline F\setminus F}} \sqrt{p(r_{\overline F\setminus F}|r_F,x)p(r_{\overline F\setminus F}|r_F,x')}
 \end{split}
\end{equation}
are small on average but not individually.

Individual smallness is precisely required to prove pairwise decoherence of histories. To see this, we note
\begin{equation}
 \begin{split}
  |&\lr{\psi(r_F,x)|\psi(r_F,x')}| = \\
  &\left|\lr{\psi(r_F,x)|\sum_{r_{F'}} \Pi(r_{F'})|\psi(r_F,x')}\right|,
 \end{split}
\end{equation}
where $F' = E\setminus F$ is the fragment complementary to $F$. Then, similar to the derivation of the probability sum rule, an application of the triangle and Cauchy-Schwarz inequality reveals that
\begin{equation}\label{eq:bound-dec-2}
 |\lr{\psi(r_F,x)|\psi(r_F,x')}| \le B_{F'}(r_F,x|r_F,x').
\end{equation}
Thus, we again see that the BC controls the amount of classicality, but this time we need a stronger condition. Again, as in Sec.~\ref{sec:classicality}, we remark that which of the two conditions---decoherence of individual histories or probability sum rule---one prefers for classicality is a matter of debate that goes beyond the scope of the present manuscript\footnote{For instance, it is strongly debated in context of the Many Worlds interpretation whether it is allowed to cut away or ignore branches of the wave function with a very small weight~\cite{SaundersEtAlBook2010}.} and in many practically relevant cases one would naturally expect both conditions to be satisfied simultaneously to a very good degree.

\subsection{Multiple events}\label{sec:multiple events}

So far we focused attention on one past event $x$ at time $s$, but the strength of the DH framework is its ability to treat multiple past events $x_\alpha$ at multiple times $t_\alpha$ using complete sets of orthogonal projection operators $\{\Pi(x_\alpha)\}_{x=1}^{m_\alpha}$ for $\alpha\in\{1,\dots,L\}$. Note that these events could refer to the same or to different subsystems, so no distinguished open system $S$ is necessary. The same construction discussed in Sec.~\ref{sec:past_events} can then be used with respect to all events:
\begin{align}
 |\Psi_t\rangle 
 &= \sum_{x_L,\dots,x_1} U_{t,t_L}\Pi(x_L) \cdots U_{t_2,t_1}\Pi(x_1)U_{t_1,t_0}|\Psi_0\rangle \nonumber \\
 &\equiv \sum_{\bs x} |\psi'(\bs x)\rangle. \label{eq:history-decomp}
\end{align}
Here, we used the group composition law of unitary time evolution, the completeness of the projectors, and we abbreviated $\bs x = (x_1,\dots,x_L)$. Finally, we write out the normalization factor of the history states explicitly, i.e., $|\psi'(\bs x)\rangle = \sqrt{p(\bs x)}|\psi(\bs x)\rangle$. As usual, $p(\bs x)$ is positive and normalized, but it does not necessarily obey the probability sum rule.

Furthermore, it is useful to introduce the notion of a subset and selection function $S$. On the one hand, $S$ is a subset of $L$ elements $\{1,\dots,L\}$, and, on the other hand, this subset is used to select elements from objects indexed by $\{1,\dots,L\}$ (without changing the notation). For instance, for $L=3$ we could choose $S=\{1,3\}\subset\{1,2,3\}$, and then we naturally define, e.g., $S(\bs x) = (x_1,x_3)$ to denote a subhistory of $\bs x = (x_1,x_2,x_3)$.

We then define past redundancy with respect to a history of $L$ events if for every $S\subset\{1,\dots,L\}$ there exist $N\gg 1$ elementary fragments $j$ with projectors $\Pi(r_j)$ at time $t$ $(\ge t_L)$ such that:

\bb{(1')} The classical MI $I_C[S(X):R_F]$ exhibits a classical \emph{redundancy plateau}, i.e., $I_C[S(X):R_F] \lesssim H[S(X)]$ for all fragments $F$ that are sufficiently large but still much smaller than the total environment: $|F|\ll N$.

This is the same definition as before with $X$ replaced by $S(X)$. A few comments are nevertheless useful. First, the elementary fragments $j$, the number $N$ of redundant elementary fragments, and the projectors $\Pi(r_j)$ can in general all depend on $S$ (which we suppressed in the notation for simplicity). What matters is that those fragments are chosen with respect to a given, natural tensor product decomposition of the environment into fragments that are deemed accessible. Note that the environment could have several sub-environments (each with its own tensor product decomposition) that redundantly encode different events, but different events could be also encoded in the same fragment. Thus, there is a considerable amount of freedom in how the fragments or projectors are chosen. Moreover, $N\gg1$ means that $N=N_S$ is much larger than the number of possible histories $S(\bs x)$, which equals $\prod_{\alpha\in S} m_\alpha$. Finally, we emphasize that the fragment size $n=|F|$ is counted relative to $N$ and not relative to the physical size of the environment.

Second, we demanded redundancy with respect to all selection functions $S$ because the existence of a redundancy plateau for $X = X_1\times\dots\times X_L$ does not by itself guarantee a redundancy plateau for any subset $S(X)$. This can be seen as a containment property to make the definition self-consistent: redundancy for $L$ events shall imply redundancy for any subset $S$ of events. Furthermore, while past redundancy implies the existence of fragments $\{F_\alpha\}$ such that $I_C[X_\alpha:R_{F_\alpha}] \lesssim H(X_\alpha)$ for a single event $X_\alpha$, it does \emph{not} imply that the redundant fragments $\{F_S\}$ with respect to $S(X)$ obey
\begin{equation}
 I_C[S(X):R_{F_S}] \rightarrow \sum_{\alpha\in S} H(X_\alpha)
\end{equation}
because different events might be correlated. Einstein's famous question ``Is the moon there when nobody looks?'' provides an example: at each time the position of the moon gets redundantly encoded into different degrees of freedom of the Universe, but once we know its initial conditions we can accurately predict its future and past evolution thanks to Newton's laws. Thus, in this case $H(X_1\dots X_L) \ll \sum_{\alpha=1}^L H(X_\alpha)$ for large $L$.

Third, all results from Sec.~\ref{sec:dependency} can be straightforwardly generalized to the present case given that for every $S$ one can find fragments $\{F\}$ such that for all ordered set of fragments $\C F$ $B_F[S(\bs x)|S(\bs x')]$ is a family of strongly decaying BCs for all $S(\bs x)\neq S(\bs x')$. Again, it is important that this holds for all $S$: a small $B_F(\bs x|\bs x')$ does not imply a small $B_F[S(\bs x)|S(\bs x')]$. Moreover, as before, it is a matter of debate whether one also demands smallness of the individual conditional BCs, see eqn~(\ref{eq:cond-BC}) with $x$ and $x'$ replaced by $S(\bs x)$ and $S(\bs x')$, to guarantee approximate decoherence of all individual histories.

More interesting for multiple events is the result that the histories themselves must be decoherent since
\begin{equation}\label{eq:bound-dec-3}
 |\lr{\psi(\bs x)|\psi(\bs x')}| \le \min_F B_F(\bs x|\bs x').
\end{equation}
As before, the proof relies solely on the triangle and Cauchy-Schwarz inequality, and therefore the bound~(\ref{eq:bound-dec-3}) might not be very tight in physically realistic situations. However, it does show that past events which get redundantly encoded must necessarily correspond to (approximately) decoherent histories. Furthermore, note that decoherence of histories satisfies a containment property: if $|\lr{\psi(\bs x)|\psi(\bs x')}|$ is small for $\bs x\neq\bs x'$, so will be $|\lr{\psi[S(\bs x)]|\psi[S(\bs x')]}|$ for any $S$.

\section{Conclusions}\label{sec:conclusions}

We introduced \emph{past redundancy}---a framework inspired by QD that conceptually generalizes it in two important ways. First, as the name suggests, it accounts for redundant records of past events, i.e., redundant past-present correlations of an evolving quantum system: this is not captured by any previous definition of redundancy. Second, past redundancy can treat multiple events at multiple times, and the definition of these events does not rely on any system-environment tensor product structure. Past redundancy also does not rely on conditional independence (like the SBS or branching form) and directly asks for classically accessible records.

\begin{figure}[t]
	\centering\includegraphics[width=0.45\textwidth,clip=true]{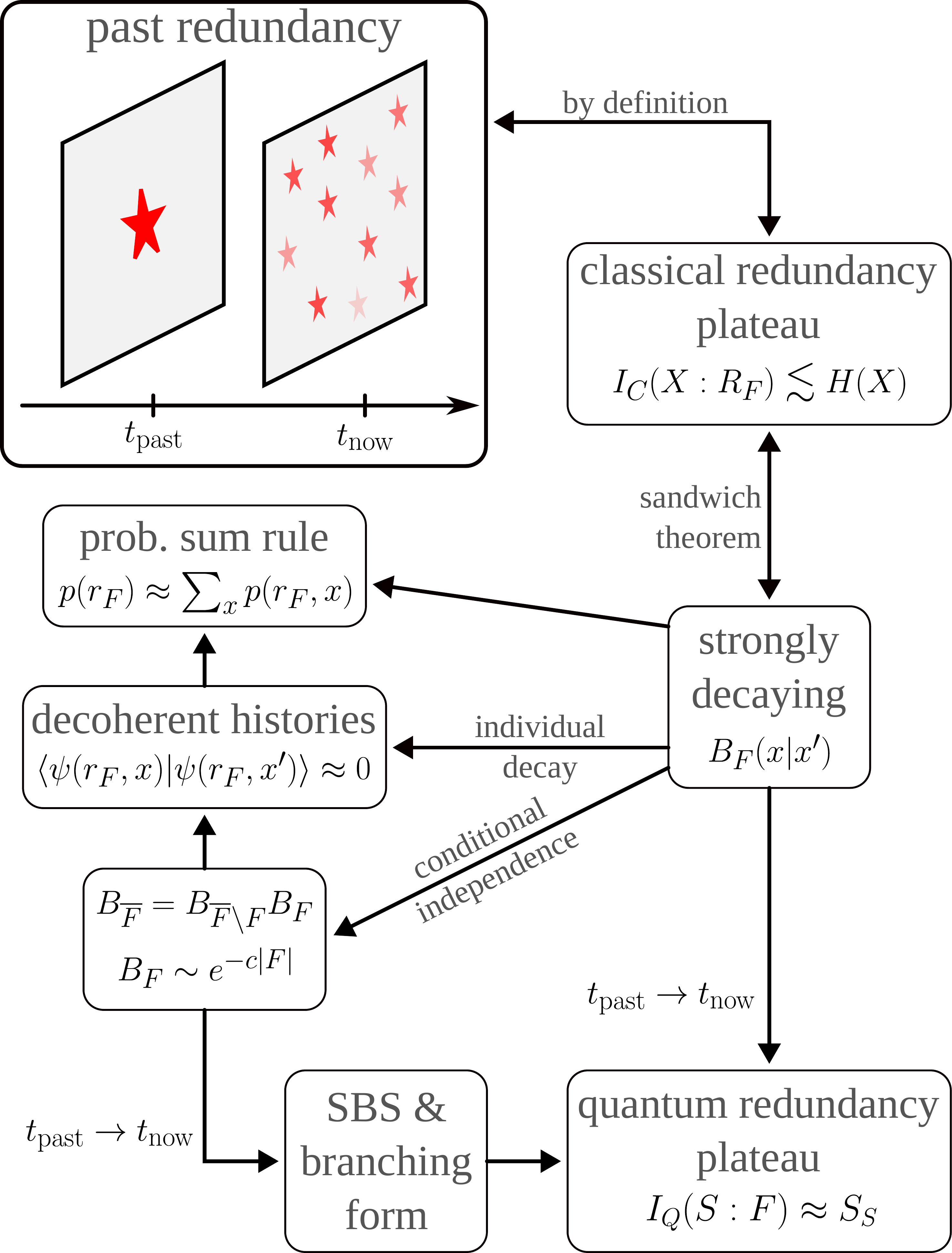}
	\caption{Overview of the results with a sketch and notation adapted to a single event $x$ in the past (indicated by a large red star) generating multiple more or less reliable records in the present (smaller and fainter stars). }
	\label{fig summary}
\end{figure}

Importantly, we showed that past redundancy keeps three key features of QD. First, single records do not need to be ideal and they can reveal little about the event---what matters is an extensive amount of redundant records. Second, the emergence of a redundancy plateau in the \emph{classical} MI puts the notion of past redundancy on a firm information-theoretic footing similar to QD---without the problem that quantum MI can contain correlations that are practically inaccessible. Third, while QD is linked to the pointer basis resulting from environmentally induced decoherence, past redundancy is linked to the existence of decoherent histories. For Markovian open quantum systems one naturally expects that these paradigms can be merged~\cite{PazZurekPRD1993, StrasbergSP2023}. Finally, past redundancy implies QD in the limit of an event happening ``now''. See Fig.~\ref{fig summary} for a summary.

What is gained from this conceptual generalization is a more realistic account about the formation of redundant records in a quantum Universe. In fact, as discussed in Sec.~\ref{sec:no past redundancy}, records are never used to infer something about the present but always about the past. Perhaps the most important consequence of this insight is then that many models which support only temporal QD support \emph{eternal} past redundancy. The emergence of past-redundancy is not restricted to unitary circuits of generalized CNOT-fan-out structure [eqn~(\ref{eq:U CNOT-fan-out})] or to fine-tuned Hamiltonians~\cite{DuruisseauTouilDeffnerEntropy2023, DoucetDeffnerPRX2024}---as we showed explicitly, it emerges for a wider class of physically relevant models. In particular, thermalization is no longer strictly opposed to the emergence of redundancy. Testing this in more complex quantum systems, such as those in Refs.~\cite{RiedelZurekZwolakNJP2012, CaoNussinovArXiv2026, SchneeweissArXiv2026}, is an interesting research direction.

We can also identify two different mechanisms responsible for the generation of past redundancy. The first relies on the mechanism of QD, which Zurek summarized as an ``environment [...] relentlessly cloning the information about the einselected pointer states''~\cite{ZurekRMP2003}. This mechanism also leads to past redundancy, but it is not the only way to generate redundancy. In particular, this QD mechanism relies on a well-defined pointer basis and it can lead to fragile records, which formally contribute to the redundancy plateau of MI while remaining practically undectable (see also the discussion about the photonic environment below). Moreover, there are important examples which clearly do not fit this ``relentless cloning of pointer states'' paradigm.

For instance, the abundant number of photons in the cosmic microwave background were released during electron-proton recombination events happening $\approx13.5$ billion years ago. It was clearly not the case that there was any relentless cloning of photons involved---in fact, energy conservation clearly forbids this (in general, conservation laws put severe restrictions on the redundant cloning of information). Another example are fossils such as dinosaur bones. Dinosaur bones encode relevant biological information about an extinct species unrelated to their pointer basis. Also here there is no relentless cloning involved, rather the opposite: the fossil recovery rate of \emph{Tyrannosaurus rex} is in between 1 per 16,000 and 1 per 80,000,000 individuals~\cite{MarshallEtAlSci2021}. Finally, consider ``destructive'' quantum measurements such as photo detectors: via a metastable cascade structure the absorption of a single photon triggers a macroscopic electric signal. At the end of the measurement there is no photon left anymore (the quantum MI between the photon system and the environment is zero for practically all times) and the pointer basis of the system is irrelevant.

All these examples point towards the fact that ``macro-events'' can create redundancy without QD, i.e., without any relentless cloning in the pointer basis. This can be anticipated from Ref.~\cite{CaoNussinovArXiv2026} and from the example of Sec.~\ref{sec:coherent state}: one needs a macroscopic superposition $|\alpha(0)\rangle + |\tm\alpha(0)\rangle$ with a large coherent state amplitude $\alpha(0)$ in order to generate redundancy and decoherent histories. For the future it is therefore worth looking at metastable quantum structures and quantum cascade systems~\cite{LeeWackerPRB2002, WoernerReimannElsaesserJPCM2004, KubisEtAlPRB2009, IakoupovEtAlPRR2026} to understand the formation of macro-events and redundancy. In fact, this research direction has gained little attention from the community working on foundational questions, even though it is believed that all macroscopic uncertainty is of quantum origin~\cite{AlbrechtPhillipsPRD2014}.

\subsection{Scattered photons as a concluding example}\label{sec:conclusions:photons}

Any discussion about the relevance of QD would be incomplete without mentioning the case of scattered photons~\cite{RiedelZurekPRL2010, RiedelZurekNJP2011, KorbiczHorodeckiHorodeckiPRL2014}. Photons are an ideal candidate for QD as they naturally constitute a non-interacting environment such that the generalized CNOT-fan-out unitary of eqn~(\ref{eq:U CNOT-fan-out}) emerges for a dynamically frozen system that scatters photons. Indeed, within the framework of past redundancy, we can immediately see that the system does not need to be dynamically frozen. As long as the rate of photon scattering is much larger than the rate of change in the objects position (which is satisfied in everyday life, e.g., when the sun illuminates a flower or a frog), different ``layers'' of the environment at a distance $d\in[x,x+\delta]$ redundantly encode the object's position a time $t \approx d/c$ earlier, leading to the perception of continuous motion since the time resolution $\delta/c$ is immeasurable small.

Is QD for scattered photons therefore an unambiguous source of redundancy, responsible for the fact why human observers in the Multiverse agree on the flower or frog they see? The calculations of Refs.~\cite{RiedelZurekPRL2010, RiedelZurekNJP2011, KorbiczHorodeckiHorodeckiPRL2014} were done in the limit $r\ll\lambda$ and $\Delta x \ll\lambda$, where $\lambda$ is the wavelength of the light, $r$ the radius of the object (a dielectric sphere) and $\Delta x = |x_1-x_2|$ the extend of the spatial superposition of the sphere. For visible light ($\lambda\approx500$~nm) we are hence talking about sizes smaller than a virus in a tiny spatial superposition. Now, the redundant information in the photons is encoded in their phase, and while the human eye is able to detect sufficiently large phase differences, the tiny phase differences caused within the approximation $r,\Delta x\ll\lambda$ are virtually undetectable. This shows that even a SBS of the quantum state, introduced to overcome the problem that quantum MI can contain inaccessible information~\cite{KorbiczHorodeckiHorodeckiPRL2014, HorodeckiKorbiczHorodeckiPRA2015}, can still encode practically inaccessible information. Human observers are in general not phase-sensitive quantum measurement devices.

What happens in the opposite limit of $r\gg\lambda$ and $\Delta x \gg\lambda$ (about the breadth of a hair or larger)? The leading order contribution to the scattered photons from such macroscopic objects then follows from geometric optics: every macroscopic object, even when localized at a single point, redundantly encodes its spatial information in a for humans easily detectable way. Let us now imagine that we like to put the object into a macroscopic spatial superposition $\Delta x = |x_1-x_2| \gg\lambda$. Owing to extremely rapid decoherence processes in everyday environments~\cite{JoosZehZPB1985, Zurek1986}, the state of the object would be described---practically for all times---by a statistical mixture $\rho(x_1)+\rho(x_2)$ instead of a coherent superposition $|\psi(x_1)\rangle + |\psi(x_2)\rangle$. We then encounter a situation in which the two mechanisms mentioned above---redundancy through QD and redundancy through macro-events---contribute to the total redundancy.

In that context it is particularly important to pay attention to the question in which degree of freedom redundancy is encoded. While QD generates redundant information in the phase of a photon, redundant information is also encoded in the color or brightness of an object. Yet, QD in the sense of a ``relentless cloning of information'' can not be related to the wavelength or intensity of light owing to energy conservation.

To conclude, that many photons contain redundant information about the object they scatter from is out of question. The mechanism for this can be QD or the macroscopicity of the object (or both). Yet, whatever the mechanism is, it is a logically distinct question whether that redundant information is relevant to humans: this is not decided by quantum MI or the SBS alone. Objectivity relies on redundancy, but it requires more than that. Past redundancy emerges as a flexible, unifying framework in which all these different mechanisms and questions can be studied. \\

{\footnotesize
\noindent\bb{AI liability statement and acknowledgements.}
I am working on DH since a few years and I am familiar with the concept of QD since many years. The motivation to combine both frameworks resulted from email exchanges with C.~Jess Riedel about the question ``how generic is QD?'' in early 2026, and from the ``branch selection problem'' that is discussed in Ref.~\cite{StrasbergEtAlArXiv2026}. When starting to think about redundancy of past events, I noticed that many models would show eternal past redundancy but only temporal QD. Eventually, this gave rise to the present manuscript. All these ideas, the main definitions and the writing of the entire manuscript are AI free. I consulted AI for specific technical questions related to information theory, quantum state discrimination and numerical implementations. In particular, most credit for the proof in Appendix~\ref{sec:app:proofs} goes to AI. Also the discussion of scattered photons, and what is visible to the human eye, profited from AI interactions.

I am very grateful to C.~Jess Riedel for critical feedback on the manuscript. Furthermore, I thank Joseph Schindler and Andreas Winter for discussions.

I am financially supported by the Ram\'on y Cajal program RYC2022-035908-I.
}

\bibliography{/home/philipp/current/references/books.bib,/home/philipp/current/references/open_systems.bib,/home/philipp/current/references/thermo.bib,/home/philipp/current/references/info_thermo.bib,/home/philipp/current/references/general_QM.bib,/home/philipp/current/references/math_phys.bib,/home/philipp/current/references/equilibration.bib,/home/philipp/current/references/time.bib,/home/philipp/current/references/cosmology.bib,/home/philipp/current/references/general_refs.bib,/home/philipp/current/references/many-body.bib,/home/philipp/current/references/non_physics.bib}

\appendix

\section{Additonal details for Sec.~\ref{sec:coherent state}}\label{sec:app:oscillators}

We first construct the optimal Helstrom measurement~\cite{BarnettCrokeAOP2009, BaeKwekJPA2015} to discriminate two coherent states $|\beta\rangle$ and $|\tm\beta\rangle$. Consider the Helstrom operator
\begin{equation}
 \Delta = \frac{|\beta\rl\beta| - |\tm\beta\rl\tm\beta|}{2}.
\end{equation}
The projectors on the positive and negative eigenspaces will give the optimal measurement, which are found to be $|\varphi_\pm\rangle = (|C_+\rangle\pm|C_-\rangle)/\sqrt{2}$ and $\mu_\pm = \pm\sqrt{1-s^2}/2$. Here, we introduced $s=\lr{\tm\beta|\beta} = e^{-2\beta^2}$ and the two \emph{orthonormal} vectors $|C_\pm\rangle \equiv (|\beta\rangle\pm|\tm\beta\rangle)/\sqrt{2(1\pm s)}$. The optimal state discrimination projectors are thus simply $P_\pm = |\varphi_\pm\rl\varphi_\pm|$, which give a success probability of
\begin{equation}
 |\lr{\pm\beta|\phi_\pm}|^2 = \frac{1+\sqrt{1-s^2}}{2} = \frac{1+\sqrt{1-e^{-4\beta^2}}}{2}.
\end{equation}
However, this reveals only the optimal measurement for a single coherent state. What is the optimal measurement for a tensor product such as $\bigotimes_{k\in F}|\pm\lambda_k\rangle$?

For this we need some linear (quantum) optics~\cite{BraunsteinVanLoockRMP2005, PetersenArXiv2016}. Namely, any linear transformation of coherent state amplitudes can be implemented in a unitary way by a passive optical network (beam splitters and phase shifters). In particular, there exists a unitary such that for any fragment $F$ of size $n$
\begin{equation}
 |\pm\lambda_1,\dots,\pm\lambda_n\rangle \mapsto |\pm\|\bs\lambda_F\|,0,\dots,0\rangle
\end{equation}
with $\bs\lambda_F = (\lambda_1,\dots,\lambda_n)$ and $\|\cdot\|$ the Euclidean norm. The optimal measurement for such a fragment $F$ thus gives a success probability
\begin{equation}
 p_F(\pm|\pm) = \frac{1+\sqrt{1-e^{-4\|\bs\lambda_F\|^2}}}{2}.
\end{equation}
The classical MI displayed in Fig.~\ref{fig PIP} is constructed from those conditional probabilities. We remark that the above strategy does \emph{not} require some globally optimized measurement on fragment $F$, but it can be implemented using a LOCC protocol~\cite{WalgateEtAlPRL2000, VirmaniEtAlPLA2001, ChenYangPRA2002}. 

\begin{figure}[tb]
	\centering\includegraphics[width=0.40\textwidth,clip=true]{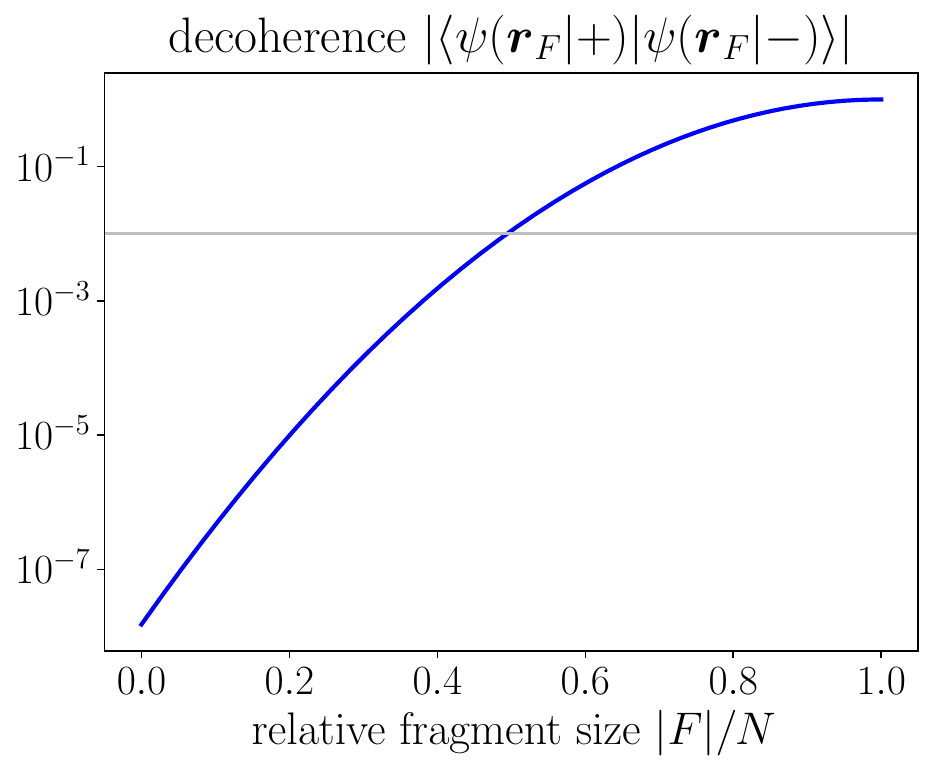}
	\caption{Plot of decoherence as a function of the relative fragment size for $\alpha(0)=3$ as used in the main text. Note the logarithmic scale. The horizontal grey line is at $0.01$. }
	\label{fig:E2:dec}
\end{figure}

To justify the interpretation of a \emph{classical} MI, we consider the DH condition. Since there are only two initial events $x\in\{\pm\}$, the only relevant overlap we need to check is $|\lr{\psi(r_F|+)|\psi(r_F|-}|$. Starting from the non-normalized history states we find
\begin{equation}
 \begin{split}
  &\lr{\psi'(r_F|+)|\psi'(r_F|-)}\\
  &= \lr{\Psi_0|\Pi(-)U_{t,0}^\dagger \Pi(r_F)U_{t,0}\Pi(+)|\Psi_0} \\
  &= \sqrt{p(+)p(-)} \lr{\psi_t(-)|\Pi(r_F)|\psi_t(+)}.
 \end{split}
\end{equation}
Here, $p(\pm) = 1/2$ are the probabilities to be in the branch $|\psi_t(\pm)\rangle = |\pm\alpha(t)\rangle\bigotimes_k|\pm\lambda_k\rangle$ prior to the measurement of fragment $F$. Thus, we obtain
\begin{equation}
 \begin{split}
  \lr{\psi'(r_F|+)|\psi'(r_F|-)} &= \sqrt{p(r_F|+)p(+)p(r_F|-)p(-)} \\
  &\times\lr{-\alpha(t)|\alpha(t)}\prod_{j\in F'}\lr{-\lambda_k(t)|\lambda_k(t)}.
 \end{split}
\end{equation}
Dividing by the probabilistic prefactors (or, equivalently, the norms of $|\psi'(r_F|\pm)\rangle$) allows us to quantify decoherence at the right relative scale:
\begin{equation}\label{eq:E2:decoherence}
 |\lr{\psi(r_F|+)|\psi(r_F|-)}| = e^{-2|\alpha(t)|^2 - 2\|\lambda_{F'}\|^2}.
\end{equation}

This result is now combined with the fact that the quadratic RWA Hamiltonian of eqn~(\ref{eq:Hamiltonian:M2}) leaves the sum of all squared coherent state amplitudes constant,
\begin{equation}
 |\alpha(t)|^2 + \|\bs\lambda_F(t)\|^2 + \|\bs\lambda_{F'}(t)\|^2 = |\alpha(0)|^2,
\end{equation}
where we used our special choice of the initial state. Thus, for long times (in regime III of Fig.~\ref{fig sys ent}) when $\alpha(t)\approx 0$ a fragment $F$ of size $n$ obeys \emph{on average} $\|\bs\lambda_F(t)\|^2 = (n/N)|\alpha(0)|^2$. Based on this we can estimate asymptotically the average strength of decoherence in eqn~(\ref{eq:E2:decoherence}), which is plotted in Fig.~\ref{fig:E2:dec}.

\section{Additonal details for Sec.~\ref{sec:Haroche}}\label{sec:app:Haroche}

We here provide a brief summary of the experimental details as far as it is needed to numerically reproduce our results, more details are given in Ref.~\cite{GuerlinEtAlNature2007, ZhouEtAlPRL2012, HarocheRMP2013} or in textbooks of quantum optics~\cite{ScullyZubairyBook1997}. 

First, the dynamics of the cavity is very well described by the standard cavity (or amplitude damping) master equation 
\begin{equation}\label{eq:ME}
 \frac{\partial}{\partial t}\rho_C(t) = \frac{1}{\tau_c} \left[a\rho_C(t)a^\dagger - \frac{1}{2}\{a^\dagger a,\rho_C(t)\}\right],
\end{equation}
where $\tau_c$ is the cavity lifetime, $a^\dagger$ the photon creation operator and $\{A,B\} = AB+BA$ the anti-commutator. Note that we work in the interaction picture. Furthermore, to numerically propagate the cavity system state in between two consecutive atom interactions, we use the exact operator-sum (or Kraus) representation~\cite{FanHuOC2009}: 
\begin{equation}
 \begin{split}\label{eq:cavity Kraus map}
  \rho_C(t+\tau) &\equiv \C E_C\rho_C(t) = \sum_{j=0}^\infty K_j \rho_C(t) K_j^\dagger, \\
  K_j &\equiv \sqrt{\frac{(1-e^{-\gamma_c \tau})^j}{j!}} e^{-\gamma_c \tau a^\dagger a/2} a^j.
 \end{split}
\end{equation}
Importantly, for a cavity with no more than $n$ photons, the infinite sum collapses to a finite sum with $n+1$ terms. 

The output state of the atoms in eqn~(\ref{eq:Haroche:interaction}) after the sudden interaction with the cavity equals for any $j$
\begin{equation}
 |\varphi(n)\rangle = \frac{1}{2}\left[\left(1-e^{-i\chi_r-i\Phi_0n}\right)|g\rangle + \left(e^{i\chi_r}+e^{-i\Phi_0n}\right)|e\rangle\right],
\end{equation}
where $|g\rangle$ ($|e\rangle$) denotes the ground (excited) state of the atom and $\chi_r$ and $\Phi_0$ are phase shifts that can be adjusted with the cavities {\tt R}$_1$ and {\tt R}$_2$ of Fig.~\ref{fig Haroche}. For our numerical implementation we choose $\Phi_0 = \pi/3$ and $\chi_r = 0$. The conditional probabilities $p(r|n)$ to detect the atom in state $r\in\{g,e\}$ given it is in state $|\varphi(n)\rangle$ are then
\begin{equation}
 p(g|0) = 0, ~~~ p(g|1) = \frac{1}{4}, ~~~ p(g|2) = \frac{3}{4}
\end{equation}
and $p(e|n) = 1-p(g|n)$. Note that many different choices for $\Phi_0$ and $\chi_r$ would give qualitatively similar results.

The implementation of the dynamics then sequentially iterates the dissipative cavity dynamics and the unitary cavity-atom interaction, keeping track of the full density matrix $\rho_{SF}(n\tau)$ of the cavity $C$ and the considered fragment $F\in\{F_1,F_2,F_3,F_4\}$, while tracing out all the other atoms, for $n\in\{1,\dots,4000\}$. Thus, there is no approximation involved in the numerical simulation.

Finally, we confirm the validity of the probability sum rule~(\ref{eq:prob sum rule}). In fact, in the present example it is actually unclear how to directly confirm the DH condition~(\ref{eq:DHC}) as the dynamics is dissipative and not unitary. However, what matters for a correct interpretation of the classical MI is the probability sum rule. Thus, we consider
\begin{equation}
 \sum_n p(r_F,n) = \mbox{tr}\{\Pi(r_F)\C E_{A_N}\C E_C\cdots\C E_{A_1}\C E_C\C D\rho_{CA}(0)\}.
\end{equation}
Here, we introduced a convenient superoperator notation: the dynamical map $\C E_C$ associated to the lossy cavity dynamics was introduced in eqn~(\ref{eq:cavity Kraus map}), the map $\C E_{A_j}\rho_{CA_j} \equiv V\rho_{CA_j}V^\dagger$ describes the unitary interaction between cavity and atom $j$ from eqn~(\ref{eq:Haroche:interaction}), and $\C D\rho \equiv \sum_n |n\rl n|\rho|n\rl n|$ is the dephasing map resulting from the initial measurement of the photon number. Now, it is possible to show that the dephasing superoperator $\C D$ commutes with $\C E_{A_j}$ and $\C E_C$ such that we can move it to the end of the trace, where it amounts to an identity operation. Thus, we get
\begin{equation}
 \begin{split}
  \sum_n p(r_F,n) &= \mbox{tr}\{\Pi(r_F)\C E_{A_N}\C E_C\cdots\C E_{A_1}\C E_C\rho_{CA}(0)\} \\
  &= p(r_F,\cancel{n}),
 \end{split}
\end{equation}
which is the probability sum rule we needed to confirm. Physically speaking, the dynamics can not turn coherences in the Fock basis into population difference that could affect the measurement statistics.

\section{Proof of the sandwich theorem}\label{sec:app:proofs}

\bb{Lower bound~(\ref{eq:bound1}).} Fix any $\bar x\neq\bar x'$ with $p(\bar x)>0$ and $p(\bar x')>0$ and introduce a new random variable $S$ whose values depend deterministically on $X$ via $s(\bar x) = s(\bar x') = 1$ and $s(x) = 0$ else wise. Since conditioning always reduces entropy we find
\begin{equation}
 \begin{split}
  H(X|R) &\ge H(X|RS) \ge q H(X|R,s=1) \\
  &= q \sum_r p_1(r) \sum_x p_1(x|r)\ln p_1(x|r),
 \end{split}
\end{equation}
where we introduced $q \equiv p(s=1) = p(x) + p(x')$ and used the shorthand notation $p_1(r) = p(r|s=1)$ and $p_1(x|r) = p(x|r,s=1)$. Next, note that by definition of $S$ we have $p_1(x|r) = 0$ if $x\notin \{\bar x,\bar x'\}$ and that, consequently, $p_1(\bar x|r) + p_1(\bar x'|r) = 1$. We thus arrive at
\begin{equation}
 H(X|R) \ge q \sum_r p_1(r) H_2[p_1(\bar x|r)]
\end{equation}
with the binary Shannon entropy $H_2$. With the reduction to the binary case we can now follow the derivation of Fuchs and van de Graaf~\cite{FuchsVanDeGraafIEEE1999}. They found (essentially as a consequence of Jensen's inequality and a clever rewriting of binary entropy)
\begin{equation}
 \begin{split}
  \sum_r &~p_1(r) H_2[p_1(x|r)] \ge \\
  &H_2\left[g\left(2\sum_r p_1(r)\sqrt{p_1(\bar x|r)p_1(\bar x'|r)}\right)\right]
 \end{split}
\end{equation}
with $g(r) = \frac{1}{2} - \frac{1}{2}\sqrt{1-r^2}$. By Bayes theorem $p_1(\bar x|r) p_1(r) = p_1(r|\bar x) p_1(\bar x)$ and analogously for $\bar x'$. Now observe that $p_1(r|\bar x) = p(r|\bar x)$ does not depend on $s$ and $p_1(\bar x) = p(x)/q$. This concludes the derivation of eqn~(\ref{eq:bound1}) after relabeling $\bar x\rightarrow x$ and $\bar x'\rightarrow x'$.

\bb{Upper bound~(\ref{eq:bound2}).} We view the different events $x\in X$ at time $s$ as different hypotheses occurring with prior probability $p(x)$. Now, given a sequence of observed records $r\in R_F$ we like to decide which hypothesis $x$ is most likely true. By Bayes' rule we have for the posterior
\begin{equation}
 p(x|r_F) = \frac{p(r_F|x)p(x)}{p(r_F)}
\end{equation}
such that the most likely hypothesis $x$ is the one that maximizes $p(x|r_F)$ for given $r_F$. This is known as the maximum a posteriori (MAP) decision rule~\cite{CoverThomasBook1991}.

We now fix $x$ and consider the error probability to falsely identify $x'\neq x$:
\begin{equation}
 \begin{split}
  P_\text{error}(x) &= \mbox{Pr}\big[\exists x'(\neq x):p(x'|r_F) > p(x|r_F)\big] \\
  &\le \sum_{x'(\neq x)} \mbox{Pr}\big[p(x'|r_F) > p(x|r_F)\big].
 \end{split}
\end{equation}
Here, we used a union bound in the second line and the probability $\mbox{Pr}$ is with respect to the observations $r_F$. Next notice that by Bayes' rule and elementary algebraic manipulations
\begin{equation}
 \mbox{Pr}\big[p(x'|r_F) > p(x|r_F)\big] = \mbox{Pr}\left[\sqrt{\frac{p(r_F|x')p(x')}{p(r_F|x)p(x)}} > 1\right].
\end{equation}
Taken together, and averaging over $x$, we thus find
\begin{equation}
 P_\text{error} \le \sum_{x\neq x'} p(x) \sum_{r_F} p(r_F|x) \chi[E(x,x')],
\end{equation}
where $\chi(E)$ is the indicator function (returning one when $E$ is true and being zero else wise) and $E(x,x') = \{r_F:p(r_F|x')p(x') > p(r_F|x)p(x)\}$. Clearly, we always have
\begin{equation}
 \chi[E(x,x')] \le \sqrt{\frac{p(r_F|x')p(x')}{p(r_F|x)p(x)}}.
\end{equation}
Thus, after some cancellations, we arrive at the bound for the error probability that decays with the same rate as the BC $B_F(x,x')$:
\begin{equation}
 \begin{split}
  P_\text{error} &\le \sum_{x\neq x'}\sqrt{p(x)p(x')}\sum_{r_F} \sqrt{p(r_F|x)p(r_F|x')} \\
  &= \sum_{x\neq x'}\sqrt{p(x)p(x')} B_F(x,x') \\
  &\le m \max_{x\neq x'} B_F(x,x').
 \end{split}
\end{equation}
Importantly, while the BC scales with the size $n$ of the fragment, the prefactor $\sum_{x\neq x'}\sqrt{p(x)p(x')}$ scales at worth with $m$ (the number of events $x$), and therefore $P_\text{error} \rightarrow0$ for $n\rightarrow\infty$. 

We can now employ Fano's inequality
\begin{equation}\label{eq:Fano}
 H(X|R_F) \le H_2(P_\text{error}) + P_\text{error}\ln(|X|-1),
\end{equation}
where $|X| = m$ is the cardinality of the set $X$. This concludes the proof.

\end{document}